\documentclass[fleqn,usenatbib]{mnras}
\usepackage[utf8]{inputenc}
\usepackage{amssymb}
\usepackage{graphicx}
\usepackage{enumerate}
\usepackage{multirow}

\newcommand{\logm}{$\log M_{\star}$}
\newcommand{\logt}{$\left<\log t\right>_{r}$}
\newcommand{\logtm}{$\left<\log t\right>_{m}$}
\newcommand{\metal}{$\left<Z/Z_{\odot}\right>$}
\newcommand{\tform}{$\left<t_{\rm{form}} \right>$}
\newcommand{\tlast}{$\left< t_{\rm{last}} \right>$}
\newcommand{\ssfr}{$\left< \rm{sSFR} \right>$}
\newcommand{\sfr}{$\left< \rm{SFR} \right>$}
\newcommand{\fburst}{$\left< f_{\rm{burst}}^{2\rm{Gyr}} \right>$}
\newcommand{\sfrts}{$\left< \gamma \right>$}
\newcommand{\rband}{$r$}

\newcommand*{\orcid}{\includegraphics[scale=0.04]{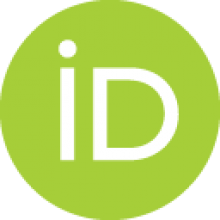}}

\title[UV upturn vs. UV weak]{UV upturn \textit{versus} UV weak galaxies: differences and similarities of their stellar populations unveiled by a de-biased sample}

\author[M.~L.~L.~Dantas, P.~R.~T.~Coelho, \& P.~Sánchez-Blázquez]{{\href{https://orcid.org/0000-0002-1178-8169}{M.~L.~L.~Dantas$^{\orcid}$}}$^{1}$\thanks{E-mail: mlldantas@protonmail.com}, \href{https://orcid.org/0000-0003-1846-4826}{P.~R.~T.~Coelho$^{\orcid}$}$^{1}$, \href{https://orcid.org/0000-0003-0651-0098}{P.~Sánchez-Blázquez$^{\orcid}$}$^{2, 3}$
\\
$^{1}$Instituto de Astronomia, Geofísica e Ciências Atmosféricas, Universidade de São Paulo, R. do Matão 1226, 05508-090, São Paulo, Brazil \\
$^{2}$Departamento de Física de la Tierra y Astrofísica, Universidad Complutense de Madrid, 28040, Madrid, Spain \\
$^{3}$IPARCOS, Facultad de CC Físicas, Universidad Complutense de Madrid, 28040, Madrid, Spain 
}
\date{Accepted XXX. Received YYY; in original form ZZZ}

\pubyear{2020}

\begin{document}
\label{firstpage}
\pagerange{\pageref{firstpage}--\pageref{lastpage}}
\maketitle

\begin{abstract}
The ultraviolet (UV) upturn is characterised by an unexpected up-rise of the UV flux in quiescent galaxies between the Lyman limit and 2500\AA. By making use of colour-colour diagrams, one can subdivide UV bright red-sequence galaxies in two groups: UV weak and upturn.
With these two groups, we propose a comparison between their stellar population properties with the goal of establishing differences and similarities between them.
We make use of propensity score matching (PSM) to mitigate potential biases between the two samples, by selecting similar objects in terms of redshift and stellar mass. Also, we take advantage of spectral energy distribution (SED) fitting results from \textsc{magphys} made available by the GAMA collaboration. The analyses are made by comparing the distributions from the SED fitting directly, as well as investigating the differences in correlations between their parameters, and finally by using principal component analysis (PCA).
We explore important differences and similarities between UV weak and upturn galaxies in terms of several parameters, such as: metallicity, age, specific star formation rate, time of last burst of star-formation, to mention a few.
Notable differences are those concerning ($g-r$) colour, metallicity, and time since last burst of star-formation: UV upturn are redder in the optical, more metallic, and their last burst of star-formation happened earlier in time. These differences suggest that UV upturn systems have shorter star-formation histories (i.e. have been evolving more passively) when compared to UV weak galaxies. Consequently, these last seem to have a higher diversity of stellar populations.
\end{abstract}



\section{Introduction} \label{sec:intro}

The formation and evolution of galaxies has only been investigated for a little over 100 years and, in that time-frame, remarkable advances have been made. \citet{Hubble1926} suggested that galaxies could be classified in two large groups: \emph{early}- and \emph{late-type} (ETGs and LTGs hereafter). The first comprises all the elliptical and lenticular systems, which are usually inhabited by aged stellar populations; whereas the latter encompasses spiral and irregular galaxies, being mostly composed by young and hot stellar populations.

It is in this context that the study of ETGs as well as the so-called ultraviolet (UV) upturn phenomenon are embedded. Since ETGs are known to be mainly constructed by aged stellar populations, their radiation peaks in the visible region of the electromagnetic spectrum, strongly weakening towards lower wavelengths -- i.e. UV. Therefore, ETGs are not expected to have any substantial emission in bluer ranges of the electromagnetic spectrum, as it would indicate hotter stellar populations (namely young stars).

Yet, some ETGs actually present a meaningful flux in the UV range of the spectrum, which has been reported for the first time by \citet{Code1969} by observing the bulge of M31; that later became the `UV upturn of elliptical galaxies' \citep[e.g.][]{OConnell1999}. Hypotheses such as residual star-formation activity in ETGs \citep[e.g.][]{Stasinska2015, Evans2018, LC&V2018} started to be considered as potential explanations \citep[e.g.][and references therein]{Yi2005, Kaviraj2007ii, Kaviraj2007i, Pipino2009, Salim2010, Bettoni2014, Davis2015, Haines2015, Sheen2016, Vazdekis2016}. Also, with the better understanding of stellar evolution, other rare stellar evolutionary phases started to be acknowledged as potential culprits for the UV upturn. Theories discussing the role of post-main-sequence stellar evolutionary phases have been proposed as early as 1971 \citep[see][]{Hills1971}. Suspects of prompting this phenomenon are blue horizontal branch (HB) stars and extreme HB (EHB) stars \citep[][and references therein]{Yi1997, Brown1998, Brown2000, Yoon2004, PengNagai2009, Donahue2010, Loubser2011, Schombert2016}, as well as the asymptotic giant branch (AGB) family of stars: post-AGB, post-early-AGB, AGB manqué \citep[e.g.][]{GreggioRenzini1990, Brown1998, Deharveng2002, Donas2007,  Han2007, Chavez2011, Rosenfield2012}, and even `regular' AGB stars \citep[see e.g.][as references to the UV excess in AGB stars]{Ortiz2019, Guerrero2020}. Besides, quiescent galaxies can also suffer from boosts in their UV emission due to binary systems in interaction, which makes them likely suspects of causing the UV upturn \citep[e.g.][]{Zhang2005, Han2007, Han2010}.

A global characteristic that appears to be a consensus in the community is the role of metallicity and its impact in the UV upturn phenomenon \citep[e.g.][]{Faber1993, Yi1997, Jeong2012, Chung2013}. \citet{Burstein1988} suggested a relation between the UV upturn and the intensity of Mg$_2$, which is supported by \citet{Boselli2005}, \citet{Donas2007}, and \citet{Bureau2011} but not backed by \citet{Loubser2011}. Moreover, some studies indicate that helium-enhanced populations might explain simultaneously the UV properties of globular clusters and the UV upturn \citep[e.g.][]{DAntona2004, Lee2005, Kaviraj2007i, Piotto2007, Peacock2011, Schiavon2012, Chung2017, Goudfrooij2018, Peacock2018}. Some studies have even found the UV upturn in NGC 6791, an old open cluster \citep[see][]{Buson2006, Buzzoni2012}.

To try to stratify this patchwork of stellar populations in galaxies, one must investigate their spectral energy distributions (SEDs). The SED of a galaxy encodes the integrated light from all the emitting astrophysical objects inhabiting therein, with special remarks on stars, stellar remnants, and dust. Therefore, breaking the encryption of SEDs is key to analyse the influence of each of those components; properties such as stellar masses, ages, metallicities, and so forth can be retrieved from this technique \citep[see][for reviews on the topic]{Walcher2011, Conroy2013}. Several previous works have attempted to analyse the potential prompters of the UV upturn; e.g. \citet{HP2013} developed a library rich in binaries and they further concluded that these are paramount to explain the phenomenon \citep{HP2014}. The analysis made by \citet{Lonoce2020} is based on absorption line features in the UV and it was applied to four massive ETGs fostering the UV upturn in the highest redshift ($z$) so far ($z \sim 1.4$); they concluded that only old stellar populations could explain such UV emission, without specifying the most likely culprits. Also, \citet{Werle2020} suggest that some galaxies presenting UV upturn can be influenced by accretion or minor mergers and their UV emission is due stellar populations younger than 1 Gyr; for their sample, this would explain the UV emission of nearly 20 per cent of the cases. On the other hand, they also believe that other UV upturn systems are mainly influenced by old and hot stellar populations (i.e. stars in the centres of planetary nabulae and white dwarfs) with an addition of EHB stars and binary systems in interaction; these account for approximately 80 per cent of their UV bright quiescent systems.

Apart from the analysis of stellar populations, several works have tried to investigate this issue in different viewpoints. For instance, \citet{Burstein1988} suggested a link between the UV upturn and velocity dispersion, which has been supported by works that have determined that, in fact, the UV upturn is more frequent in more massive ETGs \citep[e.g.][]{LeCras2016} and its dependence has been shown by \citet{Dantas2020}. Also, many works investigated whether the strength of the upturn evolves in $z$ \citep[e.g.][]{Brown2004, Rich2005, Ree2007, Boissier2018}, as well as the fraction of ETGs nesting the UV upturn \citep{Dantas2020}. Others looked for environmental dependencies, such as \citet{Yi2011} and \citet{Ali2019}, but could not provide a definitive answer to this issue, despite the predictions from helium-sedimentation theories discussed in \citet{PengNagai2009}. All in all, there are many aspects to consider while investigating such a complex problem as the UV upturn, many of which bring to light contradictory results.

Given this context, the goal of this paper is to compare overall differences and similarities among UV bright red-sequence passive/retired galaxies (RSGs, namely UV weak and upturn galaxies), according to the paradigms of \citet{Yi2011} and \citet{CidFernandes2010, CidFernandes2011}. To that end, we have applied a propensity score matching (which was applied to stellar mass, $\log M_{\star}$, and $z$) to the RSG passive/retired sample described in \citet{Dantas2020} with the aim of strongly limiting the biases between them. It is important to highlight that the main results described in \citet{Dantas2020} concern the evolution in $z$ and \logm~of the fraction of UV upturn systems; hence, this dependence is removed with the use of PSM. We make use of the value-added catalogues provided by the Galaxy Mass Assembly (GAMA) collaboration, in which results for SED fitting using \textsc{magphys} \citep{daCunha2008} are available. With these results at hand, we explore such similarities and differences, as well as their implications. To that end, we investigate their distributions directly or differences in correlations between the resulting parameters from SED fitting by making use of correlation maps \citep[e.g.][]{deSouzaCiardi2015}. Finally, a principal component analysis (PCA) is applied with the goal of verifying the results from the previous analyses.

This paper is structured as follows. In Sec. \ref{sec:dataset} we briefly describe both the dataset and methodologies used in this paper; in Sec. \ref{sec:stellarpops} we present the stellar population properties of our sample and the differences and similarities between both UV classes; in Sec. \ref{sec:mainresults} the discussion is extended; and in Sec. \ref{sec:conclusions} we present some conclusions to this study. Extra material can be found in the Appendix. For all purposes, this work made use of the standard $\Lambda$-CDM cosmological model with the following parameters: $\left\{  H_{0}, \Omega_{M}, \Omega_{\Lambda}\right\}$ = $\left\{70 \rm{km~s}^{-1} \rm{Mpc}^{-1}, 0.3, 0.7 \right\}$ -- as used by the GAMA collaboration \citep{Baldry2018}. Also, all masses herein used are in terms of solar masses ($M_{\odot}$) unless stated otherwise.

\section{Dataset \& Methodology} \label{sec:dataset}

We make use of GAMA-DR3 database \citep{Baldry2018}\footnote{Available at \url{http://www.gama-survey.org/dr3/schema/}.} aperture-matched with Sloan Digital Sky Survey 7\textsuperscript{th} data release \citep[SDSS-DR7,][]{York2000, Abazajian2009} and Galaxy Evolution Explorer GR6/plus7 Medium-depth Imaging Survey \citep[GALEX-MIS,][]{Martin2005}.
The sample herein used is the same one used in \citet[][bold rows in Table 1 therein]{Dantas2020} -- see `final sample' with WHAN emission line diagram \citep{CidFernandes2010, CidFernandes2011} classification of retired/passive systems. 

It is worth mentioning that the WHAN chart is an emission line diagnostic diagram that uses the equivalent width of H$\alpha$ -- EW(H$\alpha$) -- and the line ratio of [NII] $\lambda6583$ and H$\alpha$ (log[NII]/H$\alpha$). By using EW(H$\alpha$), it is possible to detect some of the `liny' retired/passive population of galaxies -- which cannot be done by using log[OIII $\lambda5007$]/H$\beta$ \citep*[which is the case of the Baldwin-Phillips-Terlevich -- BPT -- diagram][]{BPT1981}.

Further details about the selection criteria and treatment applied are available in \citet{Dantas2020}.

\subsection{A control sample}

This sample was de-biased by making use of propensity score matching \citep[PSM, e.g.][]{Rosenbaum1983, Ho2007}. PSM is a technique that enables us to select similar systems in terms of the properties chosen; it has only been used a few times in Astrophysical studies so far \citep[see][for other applications in Astronomy]{DeSouza2016, Trevisan2017}. By considering retired/passive UV upturn galaxies as our benchmark, we select a control sample of retired/passive UV weak systems that are the closest in terms of \logm~and $z$. The main results in \citet{Dantas2020} are the dependence of the fraction of UV upturn systems in terms of \logm~and $z$; therefore, we applied PSM considering these two parameters, to remove such dependence. To that end, we make use of the nearest-neighbour algorithm available in \textsc{scikit-learn}, a \textsc{python} package rich in several machine learning tools \citep{scikit-learn}. During the process, some UV weak galaxies -- considered UV upturn `twins' -- were accounted for multiple times; in such cases, we only considered the system once. For that reason, the final sample for UV weak systems is smaller than that of UV upturn galaxies, which is depicted in the column `After PSM' in Table \ref{tab:uvclass_psm}.

\begin{figure}
    \centering
    \includegraphics[width=\linewidth]{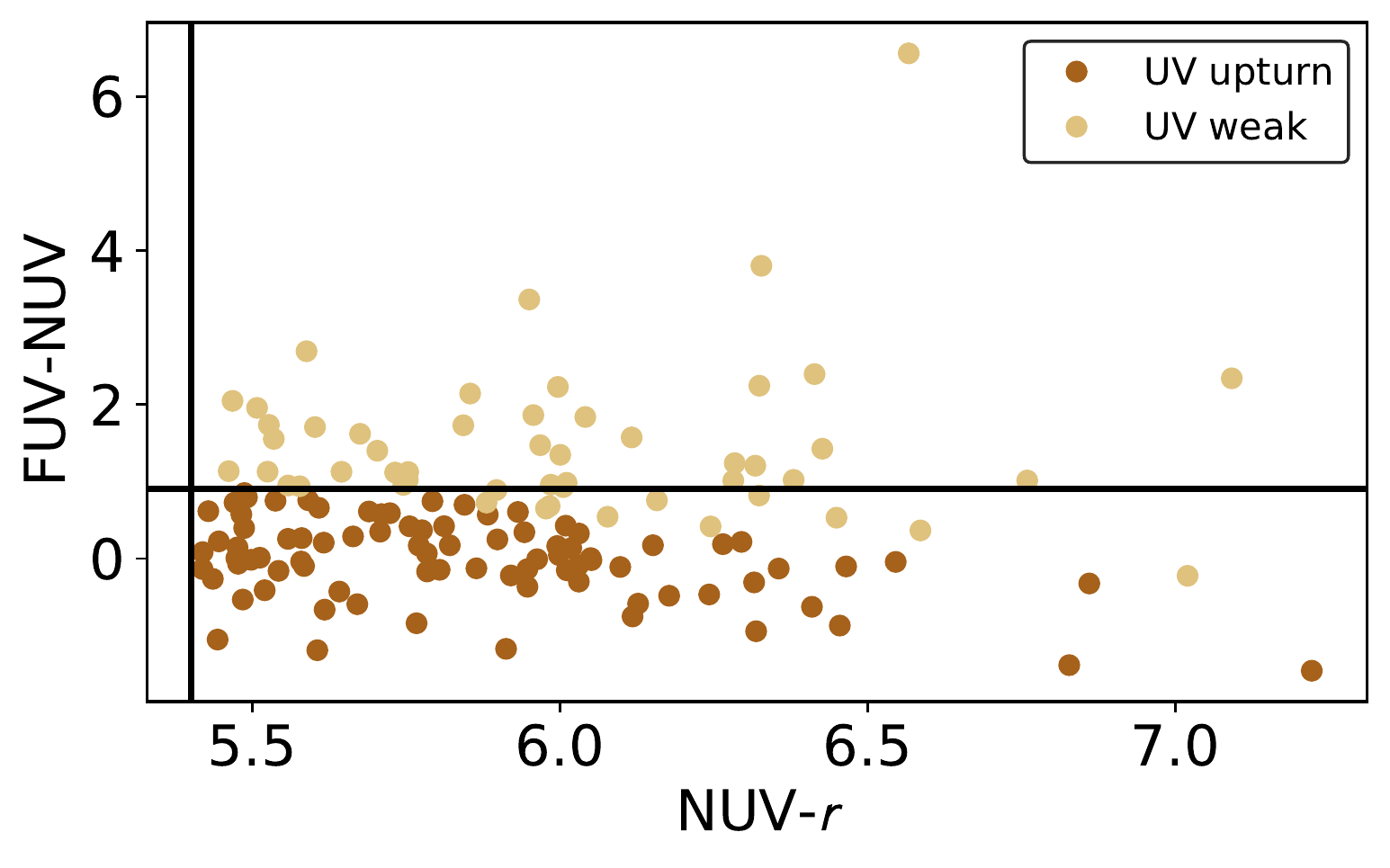}
    \caption{Colour-colour diagram for NUV-$r$ versus FUV-NUV with the classifications according to the prescription of \citet{Yi2011}. The straight black lines depict the two out of the three criteria of \citet{Yi2011}: NUV-\rband=5.4 and FUV-NUV=0.9. This image depicts the colour-colour diagram after the PSM.}
    \label{fig:yi}
\end{figure}

\begin{figure}
    \centering
    \includegraphics[width=\linewidth]{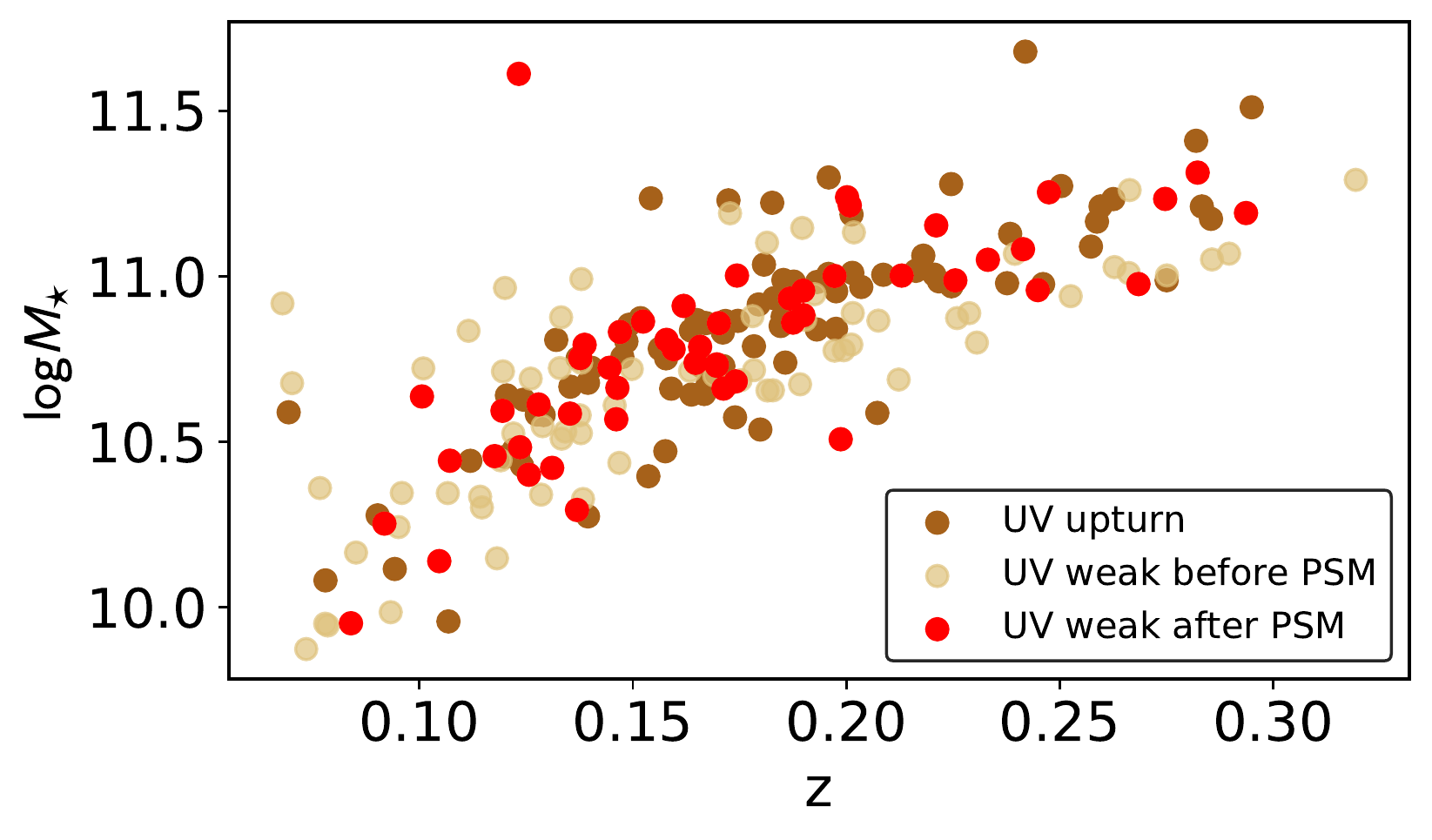}
    \caption{Scatter-plot depicting the UV upturn and weak systems. The latter is shown twice: before (in light brown) and after PSM (in strong red); the reference objects -- i.e. those nesting UV upturn -- are in dark brown.}
    \label{fig:psm_scatter}
\end{figure}

\begin{figure}
    \centering
    \includegraphics[width=0.49\linewidth]{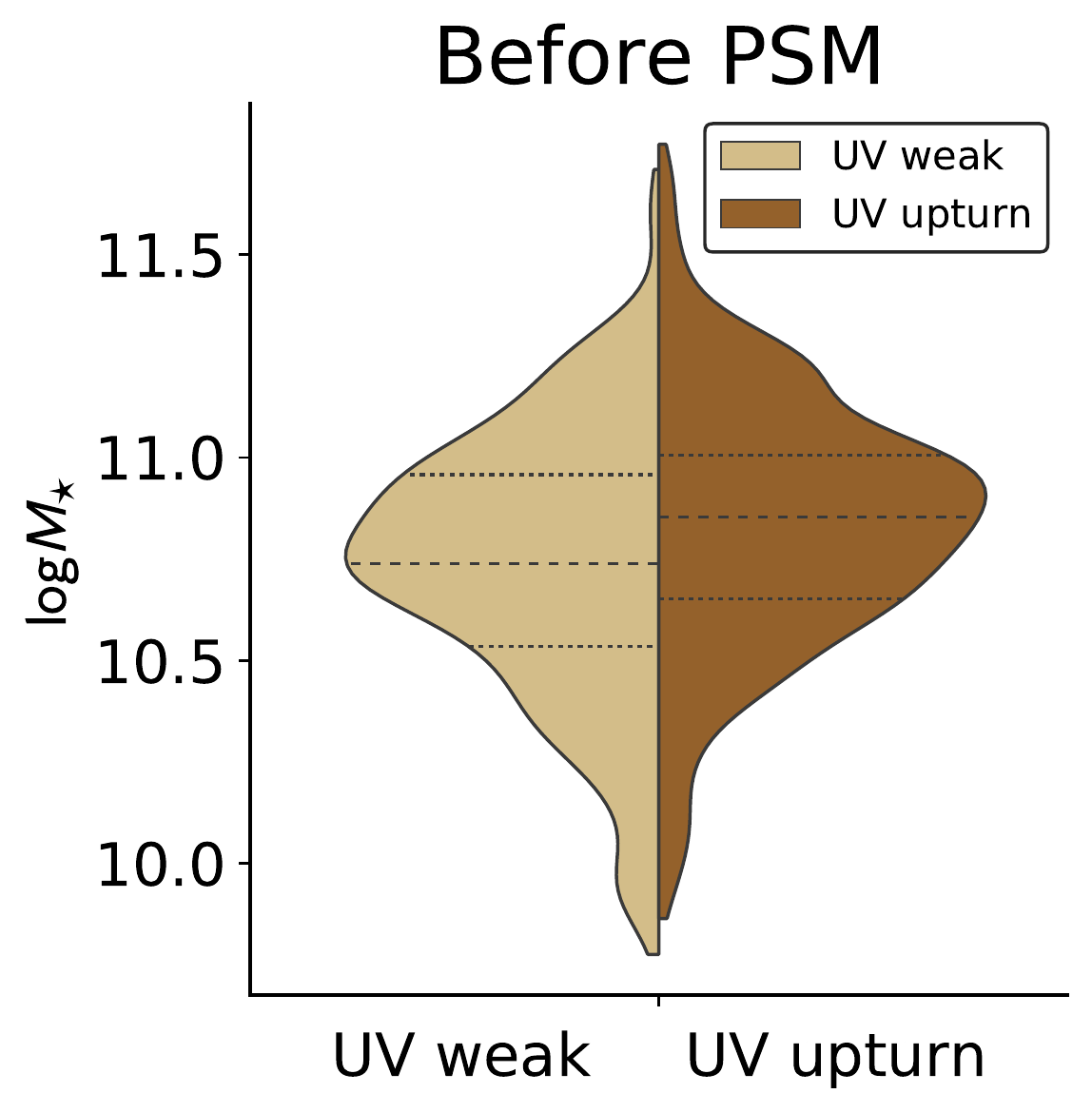}
    \includegraphics[width=0.49\linewidth]{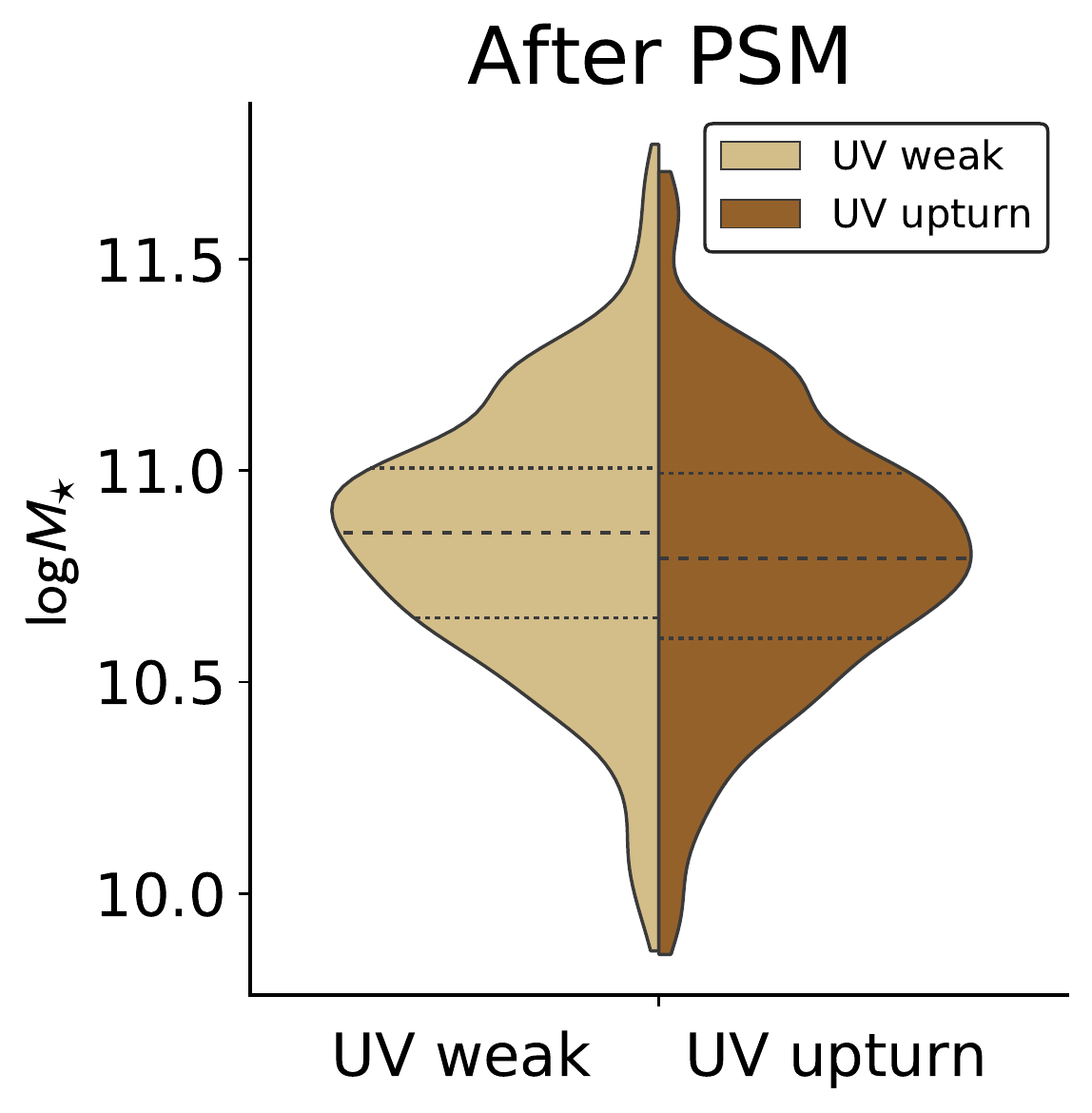}
    \includegraphics[width=0.49\linewidth, trim={0 0 0 0.92cm}, clip]{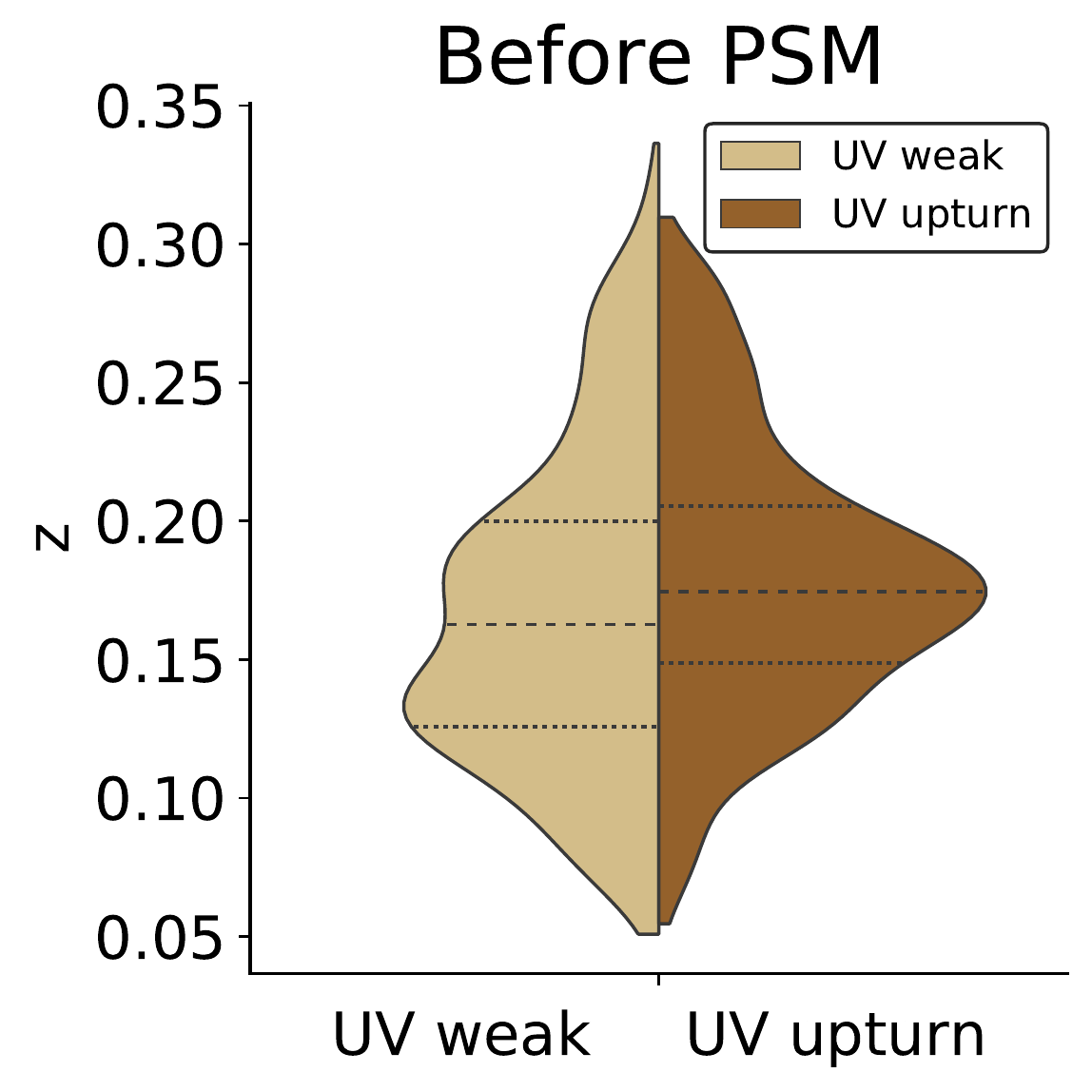}
    \includegraphics[width=0.49\linewidth, trim={0 0 0 0.92cm}, clip]{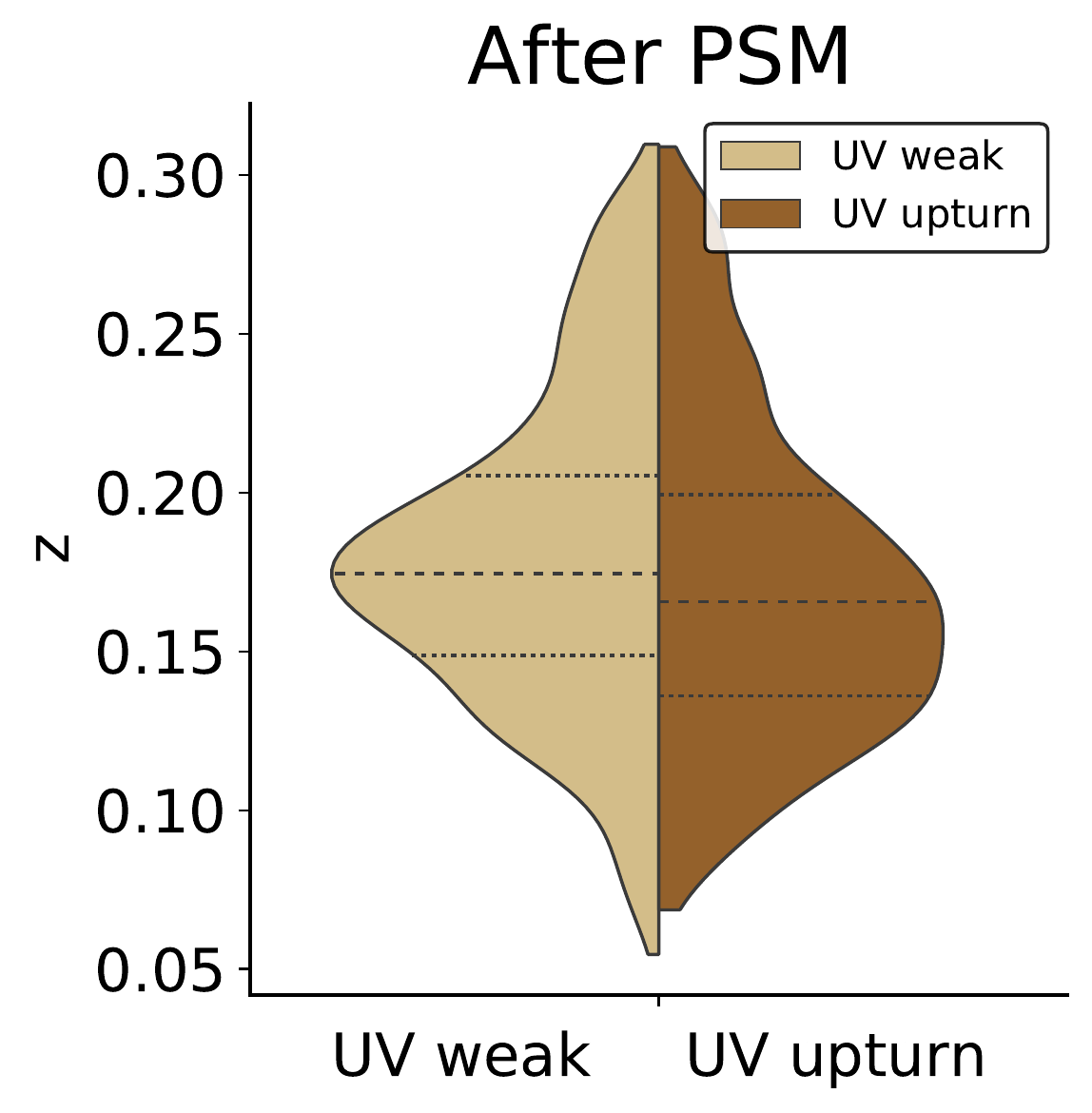}
    \caption{Violinplots displaying the distribution of \logm~(first row) and $z$ (second row) before and after applying PSM (respectively, left and right panels, as indicated by the titles) for UV weak (light brown) and UV upturn (dark brown) systems. The median and interquartile ranges are displayed in dashed and dotted lines respectively. Note that these statistical features approximate to their benchmark (those of the UV upturn); also, the shape of the distributions become more similar.}
    \label{fig:violinplots_psm}
\end{figure}

Fig. \ref{fig:yi} depicts the classification of the sample used in this work according to \citet{Yi2011}. Fig. \ref{fig:psm_scatter} displays UV upturn systems as well as UV weak galaxies before and after PSM in a chart of $z$ per \logm; the final $z$ range goes from 0.06 to 0.30. The distribution of \logm~and $z$ for both UV weak and UV upturn systems, before and after PSM, can be seen in Fig. \ref{fig:violinplots_psm}.

\begin{table}
    \centering
    \caption{Number of objects of each UV class (according to \citealt{Yi2011}) before and after PSM.}
    \begin{tabular}{l|r|r}
         Galaxy UV class  & Before PSM & After PSM\\
         \hline
         UV upturn        & 87         & 87 \\
         UV weak          & 118        & 51 \\
    \end{tabular}
    \label{tab:uvclass_psm}
\end{table}

\subsection{SED fitting}

In terms of stellar population properties, we have made use of the results from \textsc{magphys} \citep{daCunha2008} available in the homonym \texttt{MagPhys} data management unit (DMU) from GAMA-DR3 \citep{Baldry2018}. The GAMA team performed the SED fitting by making use of 21 photometric bands, spanning from the UV to the infrared (IR)/submillimetre: FUV, NUV, \emph{ugriz}, ZYJHK, W1234, PACS100/160, SPIRE 250/350/500 \citep{Driver2018} and treated them with the \textsc{lambdar} code \citep{Wright2016}. Additionally they use \citet{BC03} simple stellar population templates combined with \citet{Chabrier2003} initial mass function (IMF) and \citet{CharlotFall2000} dust models. Since \textsc{magphys} is a parametric SED fitting code, the star-formation history (SFH) is parametrised; the details on the SED fitting procedure can be found in \citet{Baldry2018, Driver2018}.

\subsubsection{Dust masses}

\begin{figure}
    \centering
    \includegraphics[width=\linewidth]{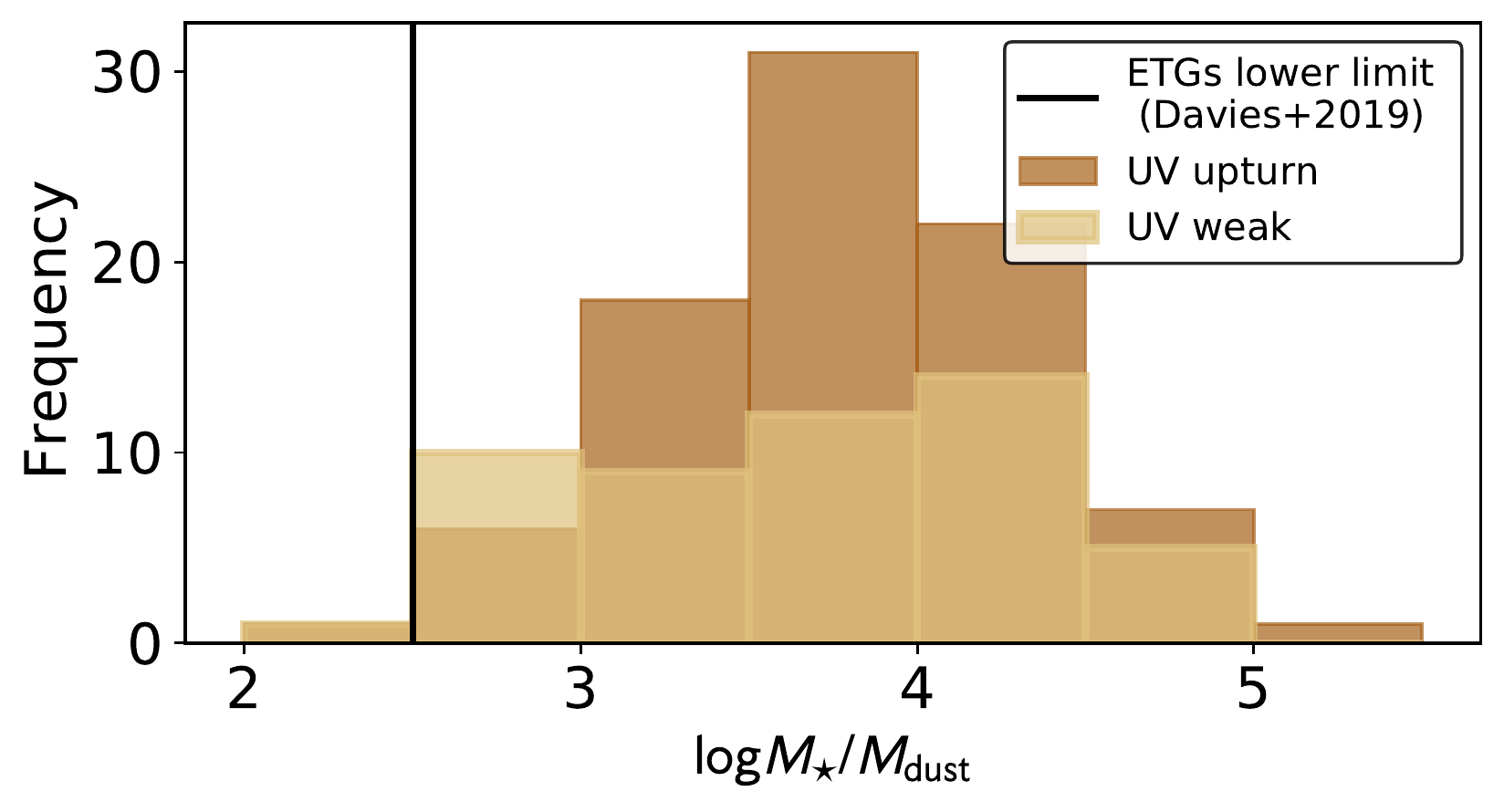}
    \caption{Distribution of the fraction of stellar and dust masses ($\log M_{\star}/M_{\rm{dust}}$) for UV upturn (dark brown) and weak systems (light brown). The vertical black line depicts the lower limit for ellipticals/lenticulars (E/S0) according to the prescription of \citet{Davies2019}.}
    \label{fig:dust}
\end{figure}

In this Section we briefly present some results for dust in our sample. We have not de-reddened the sample in terms of internal extinction albeit using SED fitting results, which may cause some contamination from reddened green-valley galaxies \citep[e.g.][]{deMeulenaer2013, Sodre2013}. To assess the impact of that, we analyse the dust mass proportion of galaxies of our sample. Fig. \ref{fig:dust} shows the ratio of \logm~and dust mass ($\log M_{\rm{dust}}$) for our sample which have been estimated through SED fitting performed by \textsc{magphys}. For comparison purposes, we added the a straight line at $\log M_{\star}/M_{\rm{dust}}=2.5$, which is the accepted lower limit value for E/S0 systems, according to \citet{Davies2019}, in cases of low dynamic interactions such as mergers. 

The results show that only one object of each UV class is heavily impacted by dust (two in total); $\log M_{\star}/M_{\rm{dust}}=2.21$ for one UV upturn system and $\log M_{\star}/M_{\rm{dust}}=2.32$ for one UV weak. Such interlopers present $\log M_{\star}/M_{\rm{dust}}$ ratio that is very close to the lower acceptable limit. Therefore, the impact of dust in our sample appears to be unimportant. Estimations of internal extinction for FUV, NUV, and $r$ bands for our sample before PSM can be seen in \citet[][Sec. A2.2]{Dantas2020}. It is possible to see therein that the reddening caused by the $r$-band ($A_r$) is small (typically $A_r \approx 0.87A_V$); also, the values for $A_g$ should be very similar to those of $A_r$. Therefore, the final reddening for the ($g-r$) colour should be next to zero, which does not cause an important impact in our analysis.

\section{Differences and similarities: a view on stellar populations} \label{sec:stellarpops}

In this section we present the main results of our analysis. For analysing correlations, we make use of Spearman's correlation rank \citep{Spearman1904}. It is worth mentioning that UV and UV-optical colours, and the 4000\AA~break are also discussed in this Section albeit not being products of SED fitting. The parameters chosen for this analysis are the following:

\begin{enumerate}[i.]
    \item D$_n$4000: 4000\AA~break;
    \item UV and UV-optical colours: FUV-NUV, FUV-$r$, and NUV-$r$;
    \item \logt: average log of age weighted by light (using $r$-band);
    \item \logtm: average log of age weighted by mass; 
    \item \metal: average metallicity in solar units;
    \item \sfr: star formation rate in the last 0.1 Gyr;
    \item \ssfr: specific star formation rate in the last 0.1 Gyr;
    \item \tform: median age of the oldest stars in the galaxy;
    \item \tlast: median time since last burst of star formation;
    \item $\left< f_{\rm{burst}} \right>$: fraction of stellar mass formed in the corresponding time-scale;
    \item \fburst: fraction of stellar mass formed in the last 2 Gyr;
    \item \sfrts: median star-formation timescale (parameter originated from the fact that \textsc{magphys} is a parametric code).
\end{enumerate}

\subsection{Direct comparison} \label{subsec:direct_comp}

\begin{figure}
    \centering
    \includegraphics[width=\linewidth]{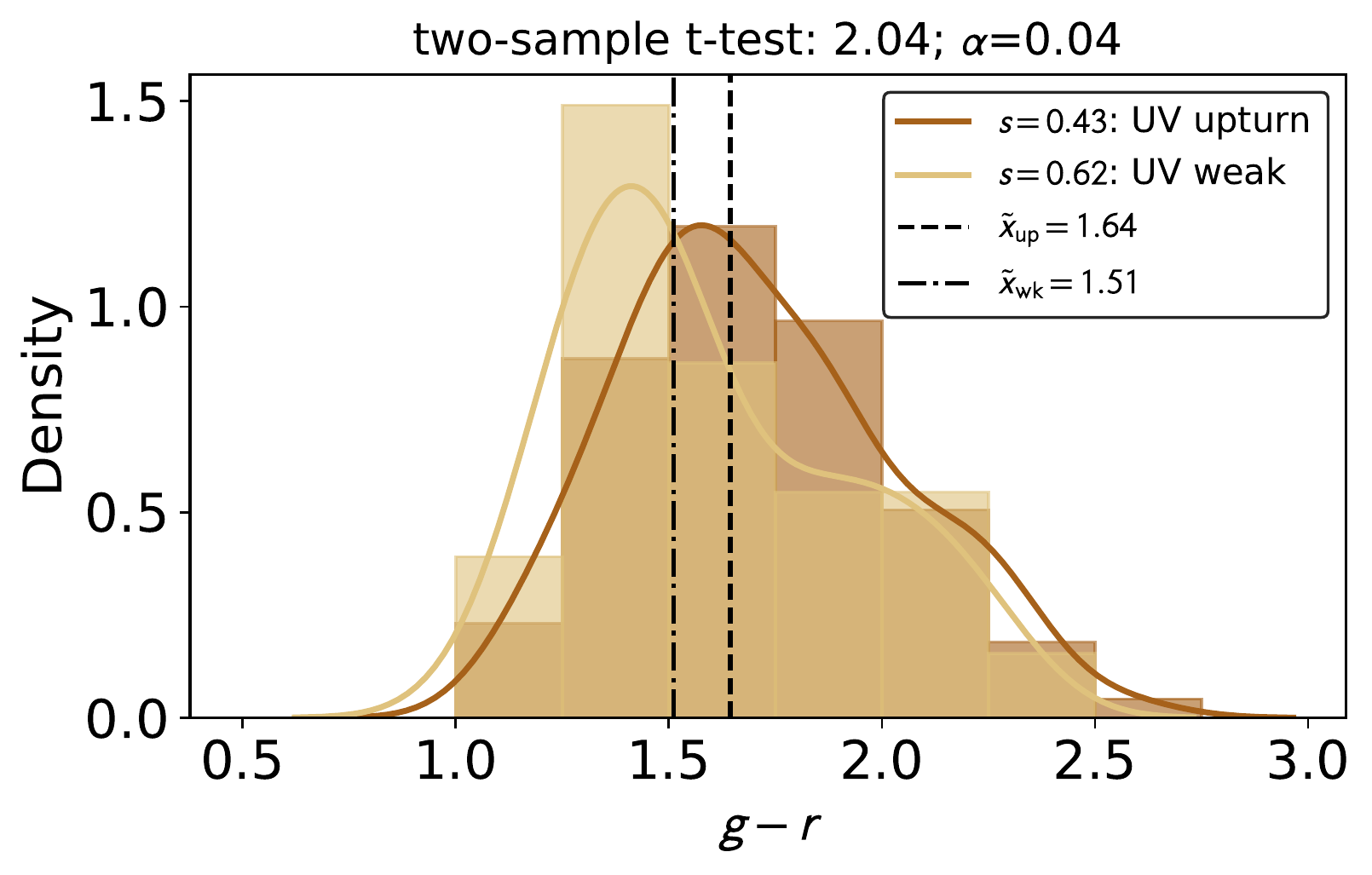}
    \caption{Distribution of ($g-r$) optical colour for UV upturn and UV weak systems, which are depicted in dark and light brown respectively. The medians ($\tilde{x}$, with indices `up' or `wk' which refers to UV upturn and UV weak respectively) are represented by black dashed and dashed-dotted lines. Also, the skewness ($s$) of both distributions are depicted in the legend box. A two-sample t-test and p-value ($\alpha$) are displayed on the title of the figure.}
    \label{fig:gr_colour}
\end{figure}

\begin{figure*}
    \centering
    \includegraphics[width=\linewidth]{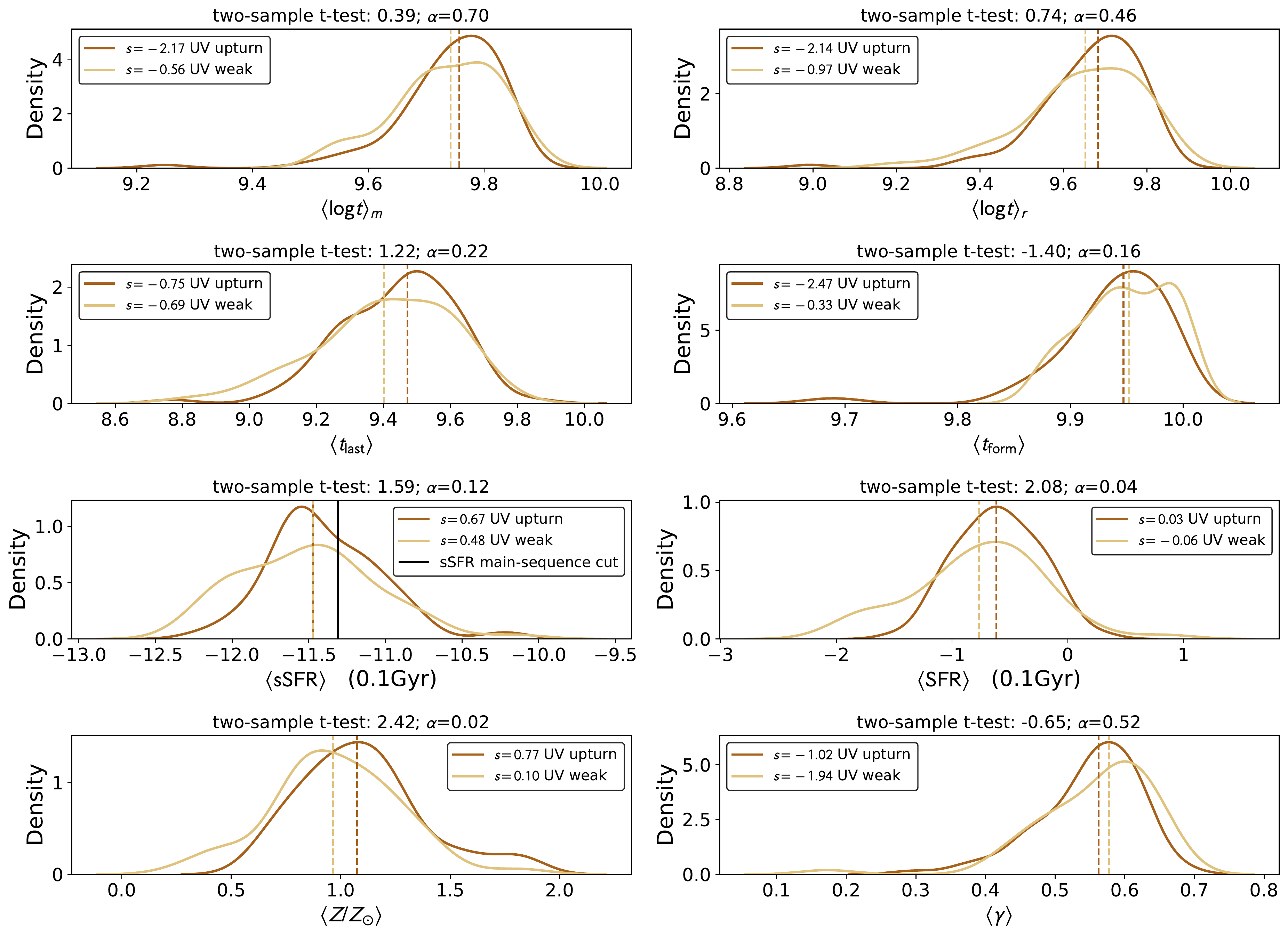}
    \caption{Distributions of six parameters which resulted from SED fitting (see text fot more details). From left to right: the first row depicts \logtm~and \logt; the second, \tlast~and \tform; the third, \sfr~and \ssfr; and, finally, the fourth, \metal~and \sfrts. UV weak and upturn galaxies are depicted in light and dark brown, respectively. The corresponding $\tilde{x}$ are in the respective colours marked in dashed lines, except for \ssfr, in which both $\tilde{x}$ coincide; in this case, the UV upturn systems are marked by a continuous line to allow for both $\tilde{x}$ to be seen overlapped. For \ssfr, a straight line at -11.31 is marked, which is considered to be a standard cut to separate passive galaxies from the star-formation main sequence \citep[see e.g.][]{LJMDavies2019}. Additionally, the skewness ($s$) of each distributions can be seen in their respective legend box. A two-sample t-test and p-value ($\alpha$) are displayed on the title of each sub-figure.}
    \label{fig:psm_directcomparison}
\end{figure*}

\begin{figure}
    \centering
    \includegraphics[width=\linewidth]{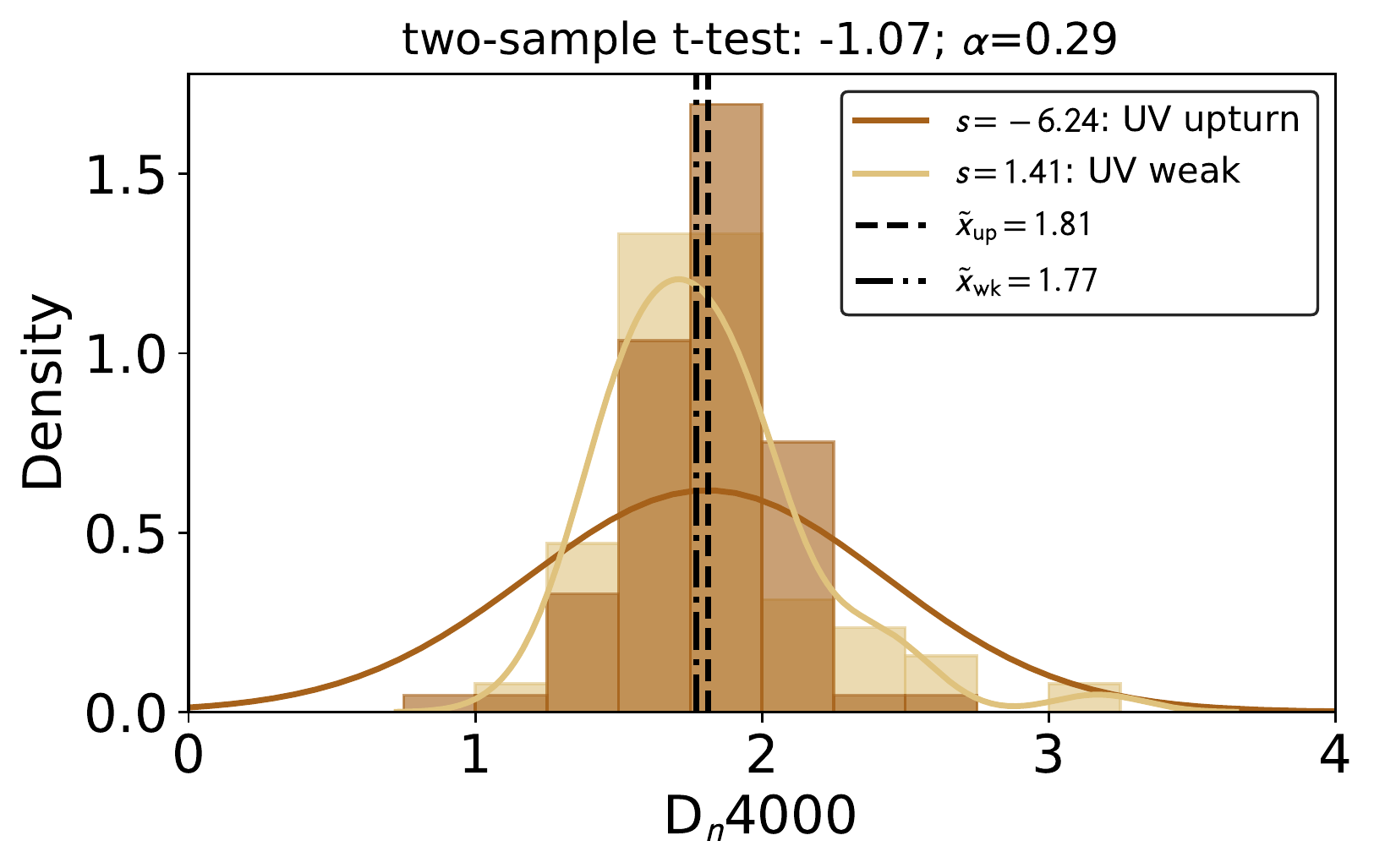}
    \caption{Distribution of D$_n$4000 for UV weak (light shade of brown) and UV upturn galaxies (dark shade of brown). Also, the skewness ($s$) of their distributions are depicted in the legend box. Their medians ($\tilde{x}$, with indices `up' or `wk' which refers to UV upturn and UV weak respectively) are depicted in the dashed and dashed-dotted lines. A two-sample t-test and p-value ($\alpha$) are displayed on the title of the figure.}
    \label{fig:dn4000}
\end{figure}

\begin{figure*}
    \centering
    \includegraphics[width=\linewidth]{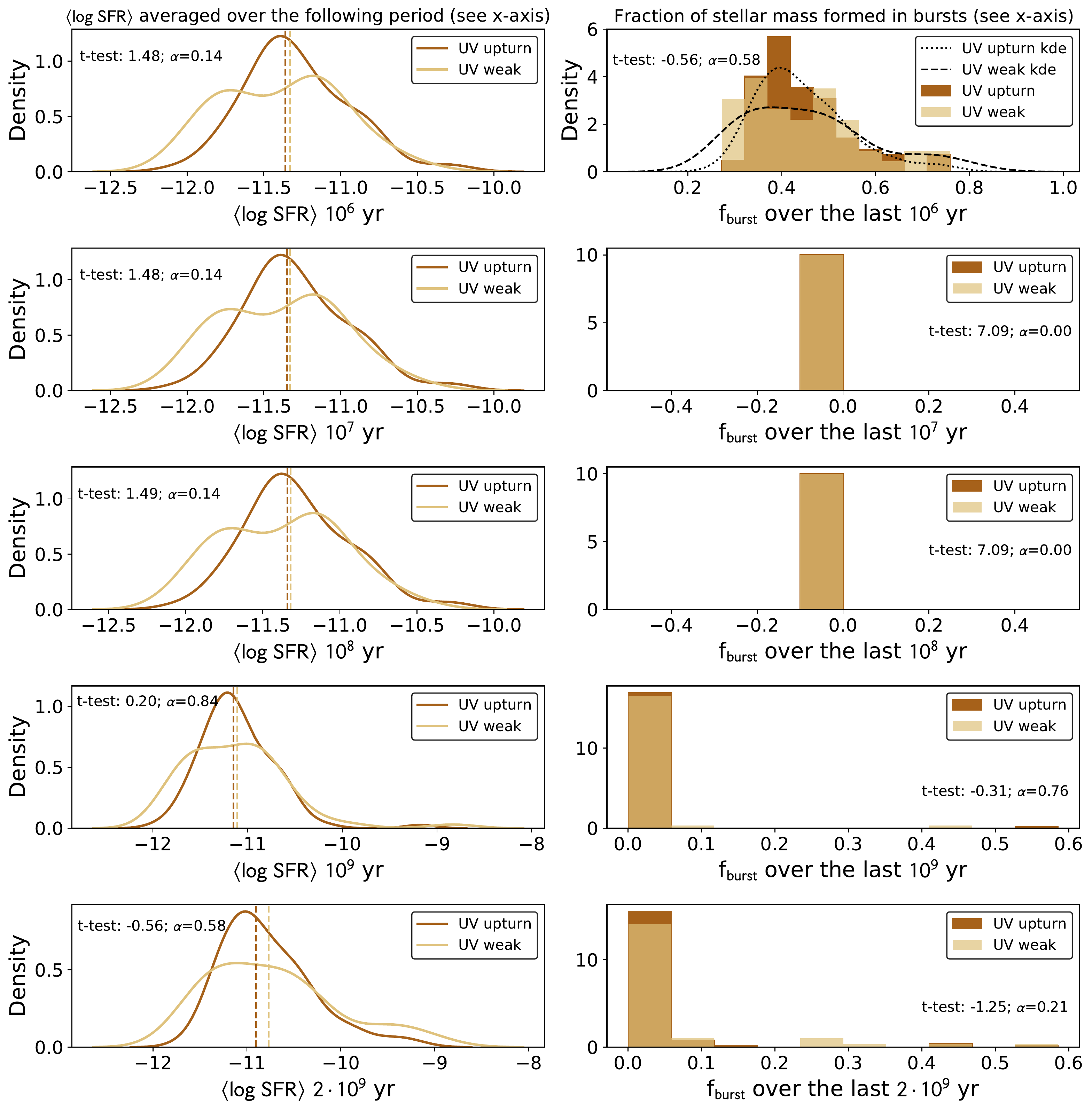}
    \caption{Averaged star-formation rate (\sfr, left column) and fraction of stellar mass formed in bursts ($\left< f_{\rm{burst}} \right>$, right column) respectively over the following periods: $10^6$, $10^7$, $10^8$, $10^9$, and $2 \times 10^9$ yr. The results herein depicted represent both UV weak and UV galaxies (in light and dark brown respectively). In all scenarios, \sfr~is very similar between both UV classes. In the first row $\left< f_{\rm{burst}} \right>$ is distributed over a range of values and a kernel density is depicted; the dotted and dashed lines depict the density for UV upturn and UV weak systems respectively. $\left< f_{\rm{burst}} \right>$ for other timescales are essentially null with some outliers; in these cases the kernel density cannot be computed. A two-sample t-test and p-value ($\alpha$) are displayed as text inside each sub-figure.}
    \label{fig:psm_sfr_frac}
\end{figure*}

In \citet{Dantas2020} we have found that UV upturn systems presented redder ($g-r$) optical colour when compared to their UV weak counterparts. Therefore, here we start our analysis revisiting this phenomenon and checking whether this behaviour repeats itself in a controlled scenario (by making use only of retired/passive UV bright RSGs after PSM). The results are depicted in Fig. \ref{fig:gr_colour}, which shows that UV weak systems are more skewed to the left than UV upturn galaxies; their medians ($\tilde{x}$) follow the trend. In fact, the behaviour seen in \citet{Dantas2020} can be reassessed in this study, even after the sample being controlled. This is an interesting finding, as the distribution of $\log M_{\star}$ is basically the same for both systems and ($g-r$) being correlated to $\log M_{\star}$ \citep[e.g. see Fig. 20,][]{Kauffmann2003_masses}. The two-sample t-test \citep{Student} is displayed on the top of Fig. \ref{fig:gr_colour}, along with the estimate of the p-value ($\alpha$ -- with usual benchmark at 0.05 for the null hypothesis). According to this criterion, the distribution of ($g-r$) is significantly different for both sets of galaxy ($\alpha<0.05$). We provide these estimates for the subsequent figures as well (Figs. \ref{fig:psm_directcomparison} to \ref{fig:psm_sfr_frac}), as some may find them useful. Nevertheless, since the use of such kind of two-sample tests and p-value can be arbitrary and controversial \citep[e.g.][]{Lin2013, Nuzzo2014, Halsey2015, Wasserstein2016}, we advise the reader to evaluate them with caution. Because of the difficulties associated with the interpretation and trustworthiness of these measures, we do not discuss them in detail throughout the text, albeit making them available. Additionally, we provide the reader with other supporting information such as $\tilde{x}$ and skewness. Skewness ($s$) is a measure of the asymmetry of a distribution; for those with longer tails to the right, $s$ is positive, negative otherwise, and null for symmetric cases \citep[for details on the topic, the reader is referred to][]{Groeneveld1984}.

Following the depiction of the ($g-r$) colour, in this Section we present the direct comparison of the distributions of the parameters from SED fitting depicted in Fig. \ref{fig:psm_directcomparison}. The sub-figures display the Gaussian kernel distribution for each parameter analysed for both types of galaxy, according to their UV class. Additionally, $s$ of each distribution is available in their respective legend box, along with the respective medians marked by the dashed straight lines.

In the first row, the Gaussian kernel density distributions of \logt~and \logtm~are displayed; they show that the stellar populations inhabiting UV upturn systems are systematically older than those in their UV weak counterparts -- \logt~medians display a difference of only $\sim 0.32$ Gyr. The analysis heretofore will consider only \logt~in order to avoid duplicate analysis.

Nonetheless, in the second row, the values of \tform~for both systems show that the oldest stellar populations of both systems have very similar ages (see the proximity of $\tilde{x}$ in dashed lines). \tlast~is higher for UV upturn than for UV weak systems, which apparently points to a gap of $\sim0.44$ Gyr\footnote{The differences in both \logt~and \tlast~were linearised to accurately depict the differences of both classes of galaxies.} between the last starbursts of the two systems; although the shape of distributions of \tlast~are similar. Yet, UV weak objects seem to have a longer tail towards low-end values.

The differences between both classes in terms of age (\logt~and \logtm)~and \tlast~are very small in terms of our current precision for age estimations \citep[e.g.][which is around 1--2 Gyr]{dotter+11}. Therefore, one should consider these differences with the according caveats. 
Yet, to probe whether the ages are actually different between both classes, we analysed the distribution of D$_n$4000 \citep[defined by][]{Bruzual1983} -- which is available in Fig. \ref{fig:dn4000}. D$_n$4000 is known to be correlated to age and metallicity and it is an empirical measurement retrieved from the spectra (observed by the GAMA survey), conveying less errors, specially for values $\gtrsim 1.3$, which is the case here \citep[e.g.][]{Kauffmann2003_masses, Mateus2006}. The analysis indicates that UV upturn galaxies have higher values of D$_n$4000 in terms of their $\tilde{x}$ and $s$ of the distribution, when compared to their UV weak counterparts. Therefore, the age and metallicity (which will be discussed further in this Section) gaps seems to be confirmed by the differences in D$_n$4000.

In the third row, \ssfr~and \sfr~are presented. The shape of the distributions for \sfr~is similar for both UV classes, except for the low-end of UV weak systems, that shows a bump that is absent in their upturn counterparts. It is interesting that according to this marker UV upturn systems have a higher rate of star-formation in the last 0.1 Gyr than compared to their weak counterparts. For \ssfr~the shapes of both systems are very different, but their $\tilde{x}$ is exactly the same at $-11.47$. A straight line in black has been added at \ssfr$=-11.31$, which is a standard cut to separate star-forming\footnote{Or in other words, the so-called `star-formation main sequence' of galaxies.} from passive systems \citep[e.g. see Fig. 2C in][]{LJMDavies2019}. It is possible to see that many of the galaxies of our sample possess \ssfr$>-11.31$; in fact 34 UV upturn and 15 UV weak systems from the post-PSM sample have \ssfr$>-11.31$, which constitutes roughly 30--40 per cent of each galaxy class. These results are not completely reliable, as the SED fitting is done by making use of \citet{BC03} models, which highly bias the UV fit towards young stellar populations. These results are not in accordance with all the measures taken to avoid star-forming effects, such as the photometric criteria according to \citet{Yi2011}, the emission lines criteria according to \citet{CidFernandes2010, CidFernandes2011}, and the amount of dust as seen in Fig. \ref{fig:dust}. Therefore, the results for \sfr~and \ssfr~are not completely trustworthy, and they must be evaluated cautiously.

\begin{table}
    \centering
    \caption{Table depicting the fraction of stellar mass formed over several timescales (from $10^6$ to $2 \times 10^9$ yr) for both UV weak and UV upturn galaxies. The descriptive statistics for each period is available below. Whenever the results are $5 \times 10^{-4}$, one may read them as null values, as are results from fluctuations in SED fitting.}
    \begin{tabular}{l|r|r}
        $\left< f_{\rm{burst}} \right>$ & UV weak & UV upturn \\
        \hline
        \hline
        \multirow{3}{*}{$10^6$ yr} & min:  0.26  & min:  0.27  \\
                                   & max:  0.80  & max:  0.76  \\
                                   & mean: 0.46  & mean: 0.44 \\
        \hline
        \multirow{3}{*}{$10^7$ yr} & min:  null & min: null\\
                                   & max:  null & max: null\\
                                   & mean: null & mean: null\\
        \hline
        \multirow{3}{*}{$10^8$ yr} & min:  null & min: null\\
                                   & max:  null & max: null\\
                                   & mean: null & mean: null\\
        \hline
        \multirow{3}{*}{$10^9$ yr} & min: null  & min: null\\
                                   & max:  0.46 & max:  0.59\\
                                   & mean: 0.01 & mean: 0.01\\
        \hline
        \multirow{3}{*}{$2 \times 10^9$ yr} & min: null & min: null\\
                                   & max:  0.53  & max:  0.59\\
                                   & mean: 0.05  & mean: 0.03\\
    \end{tabular}
    \label{tab:fburst}
\end{table}

To further investigate the results for \sfr, we compared the \sfr~and $\left< f_{\rm{burst}} \right>$ in timescales varying from $10^6$ to $2 \times 10^9$ yr; these results are available in Fig. \ref{fig:psm_sfr_frac} and Table \ref{tab:fburst}. Again, these values are low and should be evaluated cautiously. Throughout all timescales (except for $10^7$ and $10^8$ yr, in which the results are all the same and null for both types of systems), the median values of all $\left< f_{\rm{burst}} \right>$ are higher for UV weak systems when compared to their UV upturn counterparts. The distribution of $\left< f_{\rm{burst}} \right>$ for $10^6$ yr is more spread over small values, whereas for the other timescales $\left< f_{\rm{burst}} \right>$ is essentially null ($5 \times 10^{-4}$), with some outliers. This indicates that, albeit the small values of $\left< f_{\rm{burst}} \right>$, fluctuations in SED fitting indicate that UV weak systems have been forming a few more stars than those of UV upturn galaxies throughout time. This is also an inkling that \sfr~alone is not enough to explain the differences between both groups of galaxies. By using \ssfr, one mitigates some of the associating issues, but other parameters still remain important to disentangle their similarities and/or differences, such as $\left< f_{\rm{burst}} \right>$. Additionally, considering the effects of making use of stellar population templates that favour star-forming effects \citep[i.e.][]{BC03}, instead of taking into account the blue emission from rare/evolved stellar evolutionary phases, it is quite possible that these results are overestimated, specially for the last $10^6$ yr.

It is noteworthy that the analysis made by \citet{Werle2020} makes use of an updated version of \citet{BC03} models (Charlot \& Bruzual, in preparation) which accounts for many of those stellar evolutionary phases under-represented in \citet{BC03}. Therefore, their results account for a diminished influence from young stellar populations in order to explain the UV upturn phenomenon when compared to our results, specially in what concerns the values of \sfr~and \ssfr.

The forth and last row of Fig. \ref{fig:psm_directcomparison} shows the distributions of \metal~and \sfrts. The first depicts a quite different shape in the distributions of \metal~(also, note that there is an important gap in their medians and that $\alpha<0.05$); UV weak galaxies seem more skewed towards lower values than UV upturn systems. The latter shows the distributions of \sfrts, a parameter resulting from \textsc{magphys} due to its parametric characteristics; the results show that, in this case, \sfrts~is more skewed to the right for UV weak systems.

All in all, it is worth mentioning that the most important results depicted in the direct comparison analysis are the gaps in \tlast~and \metal.

\subsection{Correlation maps} \label{subsec:correlations}

\begin{figure*}
    \centering
    \includegraphics[width=0.49\linewidth]{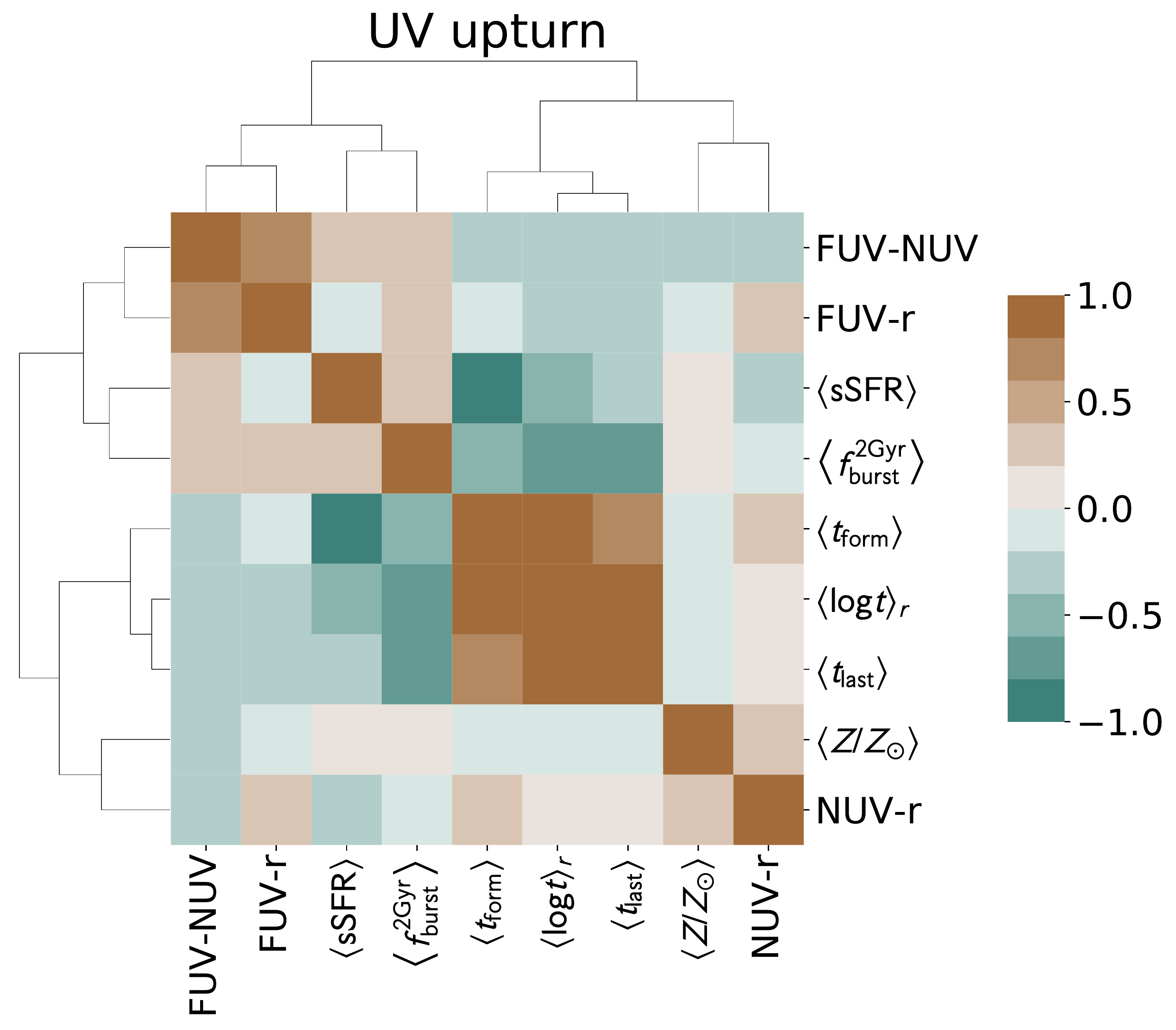}
    \includegraphics[width=0.49\linewidth]{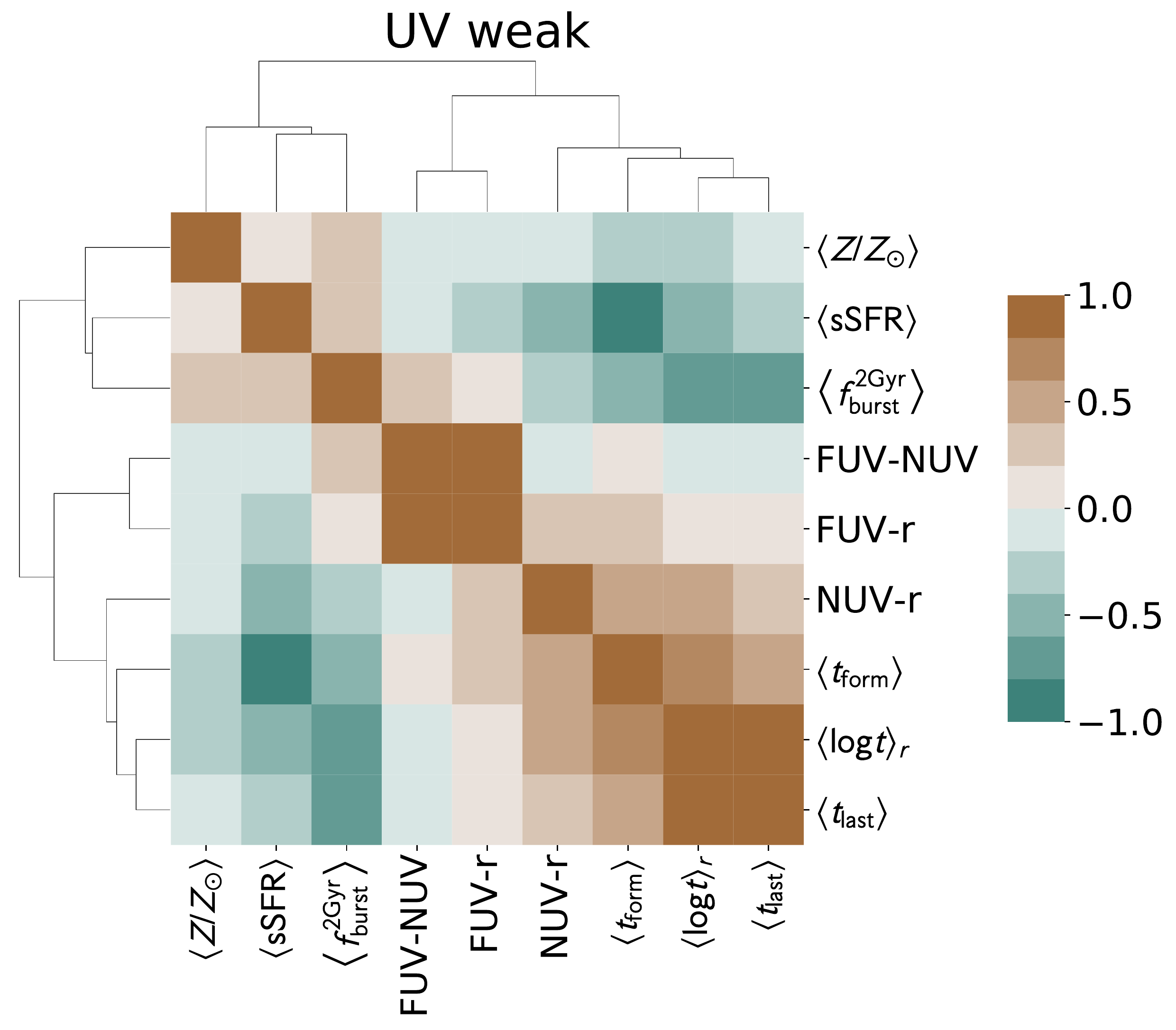}
    \caption{Cluster-maps featuring the correlation results ($\rho$, \citealt{Spearman1904}) between \textsc{magphys} resulting parameters. The UV upturn systems are in the left while the UV weak are in the right panel.}
    \label{fig:clustermaps_magphys}
\end{figure*}

Correlation maps allow us to see a myriad of relations in a very compact framework \citep[see][as a reference of data visualisation in Astronomy]{deSouzaCiardi2015}. In fact, by using heatmaps and cluster-maps, we have gained access to 36 effective values of $\rho$ for each UV class. This also means that instead of discussing each and every $\rho$ we will focus the discussion on relevant results. Additionally, with the use of dendrograms (see Fig. \ref{fig:clustermaps_magphys}), it is possible to see an hierarchical rearranging of chosen parameters and their respective correlation.

It is worth mentioning that we do not discuss the correlations for \logm, \ssfr, or \fburst~ in dedicated sections; the reasons are simple, as \logm~has been used to mitigate the biases in our sample; \ssfr~is not considered a main feature for this particular analysis; and most values of \fburst~are null. They are discussed when relevant in the following subsections.

\subsubsection{UV and UV-optical colours}

One of the main results of this analysis is the change in correlations between UV and UV-optical colours and the other parameters retrieved from the SED fitting. By simply looking at the tree last rows of Tables \ref{tab:correlations_uvwk} and \ref{tab:correlations_uvup}, it is noticeable $\rho$ changes from $\sim0$ to mild values $\rho \approx 0.2\sim0.3$. Yet, only FUV-NUV and FUV-$r$ colours are clustered together in both cluster-maps represented in Fig. \ref{fig:clustermaps_magphys}; NUV-$r$ becomes a `pivoting' parameter, switching between similar groups in each UV class (i.e. weak or upturn).

Of course, all these colours are used to perform this UV classification (Fig. \ref{fig:yi}). Therefore, we must take into account their correlations/anti-correlations, instead of their actual values. Also, the discussion between the correlations between these colours and the other parameters will not repeated in the further subsections. We further explore theses colours in Fig. \ref{fig:kde_colours}, in which FUV-NUV, FUV-$r$, and NUV-$r$ are displayed against \logt, \tform, \tlast, \metal, \ssfr, and \fburst~respectively. 2D-Gaussian kernel densities are also displayed for UV upturn (in shades of red) and UV weak systems (in shades of grey).

\paragraph{FUV-NUV:} for UV weak systems, the trend of $\rho$ between FUV-NUV and the other parameters (except other colours) is the complete lack of correlations/anti-correlations (i.e. $|\rho|<0.1$ in all cases except for \fburst). However this shifts for UV upturn galaxies: $\rho$ acquired mild values with a range of $0.22 \leq |\rho| \leq 0.33$, notably \logt, \metal, \ssfr, \tform, \tlast, and \fburst. These results are very important, as FUV-NUV is a direct measure of the strength of the upturn.

\paragraph{FUV-$r$:} the changes for this colour are more subtle. For UV upturn systems, the trend is an overall anti-correlation with the other parameters, which makes sense, since the magnitudes of the $r$-band will tend to be lower (i.e. be brighter in the optical) when compared to the FUV. However, it is worth pointing to some differences between their UV weak counterparts, with remarks on the values of $\rho$ estimated for \logt, \ssfr, \tform, and \tlast. For \ssfr, $\rho$ changes from no correlation ($\rho=-0.02$) for UV upturn to a weak anti-correlation for UV weak systems ($\rho=-0.21$); for the other three timescale parameters, they change from a weak-to-mild anti-correlations to weak-to-mild correlations -- i.e. their signal in fact changes between both types of systems --, with highlights on \tform. 

\paragraph{NUV-$r$:} this colour presents significant changes between both groups of galaxies. It is robust for \ssfr~and \tform~ though. For the other timescale parameters, \logt~and \tlast, $\rho$ is stronger for UV weak systems and non-existent for the upturn counterparts. For \metal~$\rho$ is negative and weak for UV weak systems, yet positive and mild for UV upturn counterparts. All in all, differently from FUV-NUV -- which show a clear systematic change between UV weak and upturn galaxies --, here some correlations seem more important for UV upturn systems, whereas others for weak systems.

\subsubsection{Time-scales}

The different timescales to which we refer here are \logt, \tform, and \tlast. All of them are clustered in both UV weak and upturn cluster-maps, which is an expected result, since they are not linearly independent. All the cross-correlations between them are strong, i.e. $\rho>0.6$ for all UV bright RSGs, except for \tform~\emph{versus} \tlast~for UV weak systems, in which $\rho=0.42$, which is a mild-to-strong value. Regarding the other parameters, all of these three anti-correlate with \ssfr~and \fburst~ for both UV weak and upturn galaxies, remarkably the following pairs: \tform~and \ssfr, \tlast~and \fburst, \logt~and \fburst.

\subsubsection{\metal}
The values of $\rho$ are robust in the lack of correlation with \ssfr~and \tlast~(i.e. for both groups of galaxies). Yet, differences are remarkable when looking at \logt, \tform, and \fburst: $\rho\approx0$ for UV upturn systems, but are mild ($0.2 < \rho < 0.4$) for their weak counterparts. It is remarkable that these trends are not the same for all time-scales herein used, being non-existent for UV upturn galaxies, but only changing to mild values \logt~and \tform~-- remaining robust and absent for \tlast. These results are easily seen in Fig. \ref{fig:heatmaps_magphys} due to the order of the parameters that remain the same -- differently from those of Fig. \ref{fig:clustermaps_magphys} that change due to the clustering features of the top and lateral dendrograms.

\begin{figure*}
    \centering
    \includegraphics[width=0.96\linewidth]{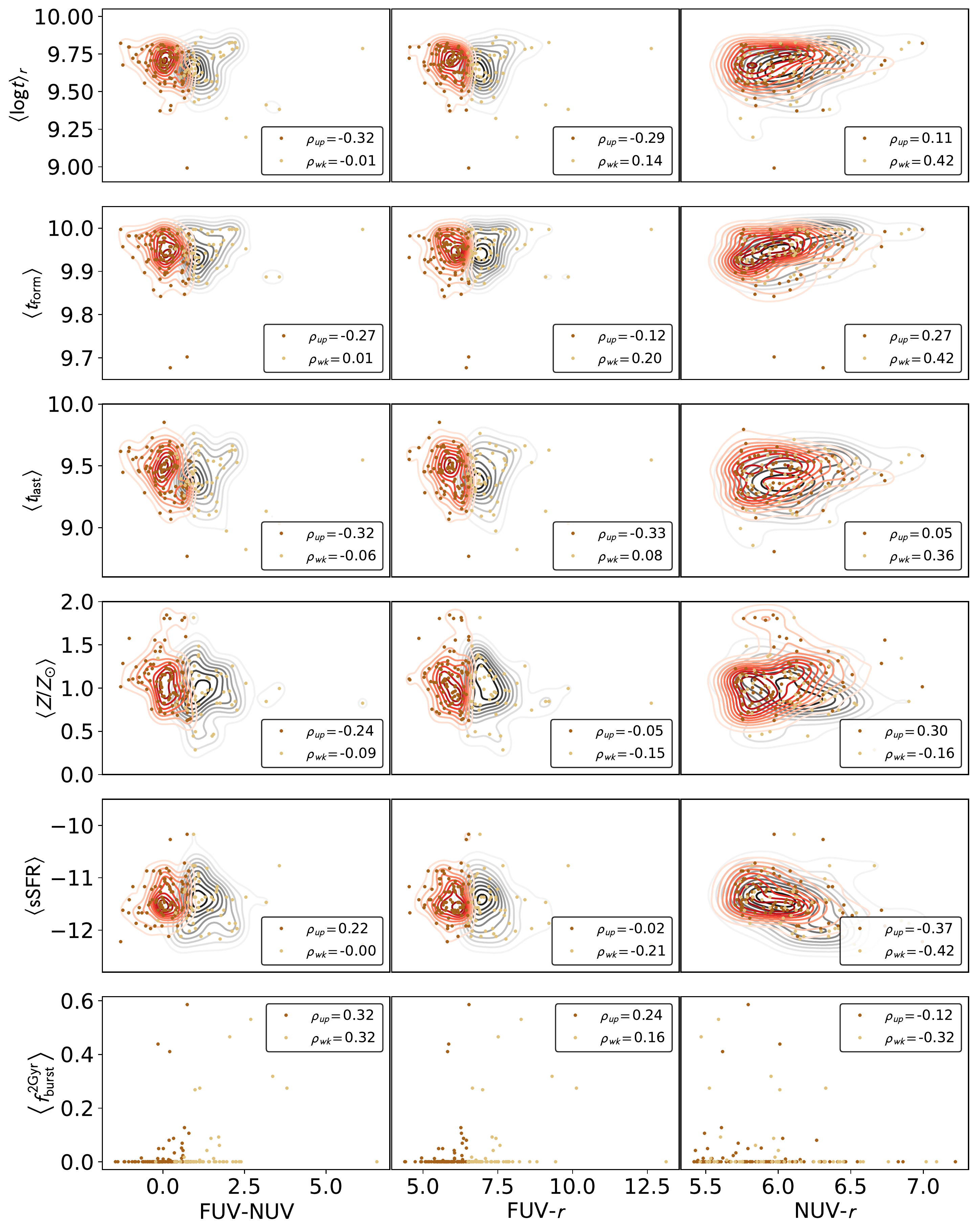}
    \caption{UV and UV-optical colours (FUV-NUV, FUV-\rband, and NUV-\rband~respectively) against the following physical parameters (from top to bottom): \logt, \tform, \tlast, \metal, \ssfr, and \fburst. UV weak and UV upturn systems are respectively depicted by light and dark brown markers, as well as grey and red 2D-Gaussian kernel density curves -- with the exception of the last row, which has many values at zero, preventing the kernel density to be estimated. The labels depict the corresponding values of $\rho$ for UV upturn ($\rho_{\rm{up}}$) and UV weak galaxies ($\rho_{\rm{wk}}$).}
    \label{fig:kde_colours}
\end{figure*}

\subsection{Principal component analysis}

In this Section, we explore the differences and similarities between UV weak and UV upturn galaxies by making use of PCA. It is a technique suitable for analysing high-dimensional datasets. The concept behind PCA is the projection of these datasets in lower dimensions by minimising the loss of information as much as possible \citep[e.g.][]{Abdi2010, Jollife2016}. It has been widely used by the astronomical community in the analyses of several different problems \citep[e.g.][]{JeesonDaniel2011, Chen2012, DeSouza2014, Pace2019}.

PCA can be used in order to verify the differences and/or similarities of the two samples of galaxies. Considering that the total variance of each type of galaxy is caused by the combined variance of  the chosen parameters, it is possible to assess which parameters contribute the most for the total variance for UV weak and UV upturn systems. By making use of PCA, one expects to find major differences up-front, determining which parameters set the differences between the two groups of galaxies. Otherwise, if these parameters are somewhat the same, this is a strong indication that these two sets of galaxies are in fact very similar. This approach is also very useful to probe the results found in Secs. \ref{subsec:direct_comp} and \ref{subsec:correlations}.

In order to obtain these results, we made use of the PCA functionality available in the \textsc{skcikit-learn} \textsc{python} package \citep{scikit-learn}. The results used are the outputs of the attribute \texttt{components\_}, which can be positive or negative; we make use of the absolute values to facilitate its visualisation\footnote{By making use of the absolute values, no loss of information happens, as the signal estimated for \texttt{components\_} is roughly a choice of methodology.}. It is worth mentioning that, since PCA is sensitive to scale, we have standardised all the variables before applying it \citep[which is a common practice to PCA users; see Sec. 2C in][]{Jollife2016}.

\begin{figure}
    \centering
    \includegraphics[width=0.98\linewidth]{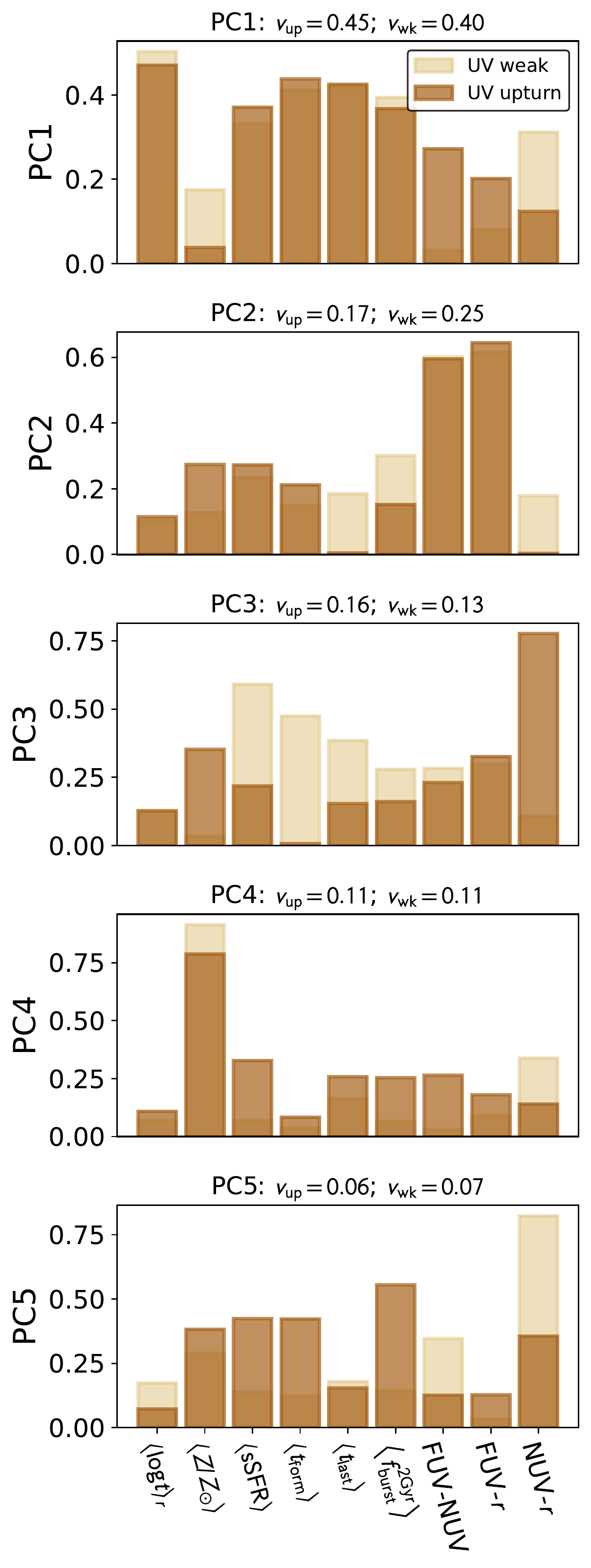}
    \caption{Bar-plots with each PC resulting from PCA analysis for both UV weak and upturn galaxies. The titles display the percentage of the contribution of each PC to the total variance for both systems ($v_{\rm{up}}$ for UV upturn and $v_{\rm{wk}}$ for UV weak).}
    \label{fig:pca}
\end{figure}

Fig. \ref{fig:pca} displays the results of the PCA analysis with 5 principal components (PCs), which describe the variance of over 95 per cent of both samples (95.1 per cent for UV upturn and 96 per cent for UV weak). The partial contributions of each PC can be seen in the titles of each sub-plot of the same image.

The results show that there are no outstanding differences between the two types of systems in the first two PCs. The main contributors for the variance of PC1 and PC2 are similar, i.e. \logt, \tform, and \tlast~are the three main parameters for both UV weak and upturn galaxies for the variance of PC1.
The contribution of \logt~for the variance are similar for both types of galaxies in PC1 (and this continues to be true throughout all the other PCs), which indicates that the differences between ages might not be significant. In a `secondary' level of PC1, FUV-NUV is important for UV upturn systems, whereas NUV-\rband~plays an important role for the variance of their weak counterparts. This is in agreement with the analysis previously made for the UV and UV-optical colours (differences in correlation parameters between both groups of galaxies, and their dispersion seen in Fig. \ref{fig:kde_colours}). Also, the difference in variance for \metal~is eye-catching; this is also in accordance to the results seen in Sec. \ref{subsec:direct_comp}.

A similar trend occurs in PC2, in which FUV-\rband~and FUV-NUV are the main contributors for the variance for both UV weak and upturn systems. In fact, it is only in PC3 that these main contributors change a lot: NUV-\rband~become very important for the variance in UV upturn galaxies, whereas \ssfr~and \tform~matter the most for UV weak systems. It is worth mentioning that in PC3 there is an important gap for \metal, which is important for UV upturn systems and nearly nonexistent for UV weak galaxies. PC4 shows, again, that the contributions seem similar for both groups of galaxies, with an important contribution of \metal~for both. Finally, PC5 depicts differences between the parameters that mostly contribute to the variance among both types of galaxies, being NUV-\rband~the most important for UV weak galaxies, whereas \fburst~for UV upturn systems. 

Finally, the analysis made with PCA serves as an independent test with the goal of confirming the analyses made in previous Sections. The results indicate that the main components responsible for the variance of both types of galaxies are basically the same, which is expected as these systems are in fact similar. They are both in the red-sequence, are classified as retired/passive systems according to their emission line features, possess the same range of \logm~and $z$, and so forth. The differences between the samples are more subtle, and can only be seen in a secondary level of PC1 (namely \metal~, FUV-NUV, and NUV-$r$), and then at other less important components (such as PC3 -- see \tform, \tlast, and \ssfr). 
\section{Main results} \label{sec:mainresults}

In this Section, we briefly expand the discussion of the main results presented in Sec. \ref{sec:stellarpops}.

In terms of the direct comparison of the distribution of the parameters of UV weak and UV upturn galaxies, an overall characteristic that is very important to observe is that $s$ for UV upturn systems is systematically higher than their UV weak counterparts. This behaviour is true for cases, either when $s$ is positive or negative; the only exception is for \sfr, when the opposite occurs, but it then disappears when it is pondered by \logm~(i.e. \ssfr). Yet, when further investigating the distributions of \sfr~combined with $\left< f_{\rm{burst}} \right>$ across several timescales, it is possible to see that UV weak systems have systematically higher $\left< f_{\rm{burst}} \right>$, when compared to UV upturn galaxies. 

However, it is important to mention that the simple stellar population templates of \citet{BC03} do not include many of the rare evolved stellar stellar evolutionary phases known to be UV bright (e.g. HB, and EHB stars), as discussed in Sec. \ref{sec:intro}. This is an important caveat, which makes that most of the UV emission be fitted by making use of young stellar populations. This is certainly impacting on the estimates of some parameters, including \sfr~and \ssfr. Yet, by making use of photometric bands in the IR, we believe that major effects of young stellar populations have been controlled for. For more accurate results, this analysis needs to take into account templates that consider these types of stellar populations.

Important results are those regarding ($g-r$), \metal, ages, and \tlast. UV upturn systems appear to be older than their weak counterparts for both \logt~and \logtm; also, \tlast~presents a median gap of $\sim0.44$ Gyr between both systems, which point to a longer period of passiveness for UV upturn galaxies. Combined with the results for \metal~-- which show lower $\tilde{x}$ values and considerably lower $s$ -- and D$_n$4000, it seems that UV upturn systems have shorter star-formation histories when compared to their UV weak counterparts \citep[these results are in agreement with][]{Werle2020}. There are two possible scenarios that could explain this difference in star-formation histories:

\begin{enumerate}[i.]
    \item UV upturn systems could have simply completed their star-formation processes earlier in time, considering that they have been through similar processes and interactions throughout their evolution;
    
    \item or UV weak galaxies, for some reason, have been more efficient in accreting metal-poor gas from the intergalactic medium, bringing their median metallicities to lower values and forming a few more stars than their UV upturn counterparts.
\end{enumerate}

Finally, it is worth mentioning that generally the parameters of UV upturn systems seem to be less disperse than those of UV weak galaxies (e.g. see Fig. \ref{fig:kde_colours}); in other words, they occupy a smaller range of values than compared to those of UV weak systems. These differences are supported by the correlations and PCA analyses (with highlights to PC1), which seems to be in agreement with the hypothesis that UV upturn systems appear to be evolving more passively when compared to UV weak galaxies. In other words, bigger dispersion of UV weak systems maybe linked to a higher diversity of stellar populations.

\section{Summary \& conclusions} \label{sec:conclusions}

We have analysed 87 UV upturn and 51 UV weak galaxies observed by the GAMA collaboration with aperture-matched data from SDSS-DR7 and GALEX GR6/plus7. To produce the aforementioned sample, we have made use of the UV bright RSGs classified as retired/passive according to the WHAN diagram \citep[which are described in further detail in][]{Dantas2020} and applied PSM, mitigating the effects of the confounding variables ($\log M_{\star}$ and $z$). From this sample, we made use of the SED fitting results from \textsc{magphys} available in the value-added catalogues from GAMA-DR3, which were estimated by making use of 21 band-passes which span from the UV to the far-IR/submillimitre. With those at hand, we have analysed their similarities and differences by comparing the distributions of their SED fitting parameters, investigating their correlations, and checking their principal components through principal component analysis. Our conclusions are as follows.

\begin{enumerate}[1.]
    
    \item First and foremost, both UV weak and UV upturn galaxies, according to this study, are very similar. This can be easily seen in the PCA analysis as a whole, but also in the direct comparison of their properties. 
    
    \item Their differences are subtle and appear in a finer analysis, such as seen for instance by the secondary parameters for PC1 (FUV-NUV, NUV-\rband, and \metal).
    
    \item UV upturn systems are redder in the ($g-r$) optical colour when compared to their UV weak counterparts in the same conditions: both being classified as retired/passive systems and occupying the same range of $\log M_{\star}$ and $z$. These results are in agreement to the previous analysis made in \citet{Dantas2020} and are consistent with differences in age and metallicity.

    \item Overall UV weak systems reach larger ranges (with longer tails over smaller values and lower skewness) generally over the SED fitting output parameters, notably age (\logt~and \logtm), \tlast, and \metal. This is a hint that suggests that UV weak systems may have a larger assortment of stellar populations.
    
    \item Stellar populations in UV upturn galaxies seem to be systematically older than their UV weak counterparts (their $\tilde{x}$ shows a gap of $\sim0.32$ Gyr), but the associated errors are too high to enable any strong assumption (which is confirmed by the PCA analysis). Also, the age of the oldest stars in both systems (\tform) is basically the same. Additionally, their time since last burst of star-formation (\tlast) has a median gap of $\sim0.44$ Gyr, indicating a potential longer passiveness of UV upturn systems. Yet, the contribution of total variance of \tlast~in PC2 and PC3 in the PCA analysis shows different contributions in for these systems; there is a higher contribution to total variance for UV weak systems, which supports the idea of higher passiveness of UV upturn galaxies. These must be interpreted with the according caveats regarding our current limitations in terms of age determination, and the use of \citet{BC03} models, which do not properly account for the UV emission from rare/evolved stellar evolutionary phases. Yet, other major issues regarding the overfitting of young stellar populations have been mitigated by the use of photometric bands in the IR.
    
    \item An additional analysis made with D$_n$4000 shows another gap between these two types of galaxies, which may help corroborate the the results for age and metallicity.
    
    \item UV weak galaxies have been showing slightly higher mean values of $\left< f_{\rm{burst}} \right>$ throughout several timescales (from $10^6$ to $2 \times 10^9$ yr). This indicates that UV weak systems have been to some extent more efficient in forming new stars, albeit the closeness of \sfr~throughout the same timescales between UV weak and UV upturn systems. This is another indication of higher passiveness of UV upturn galaxies.
    
    \item Correlation parameters between UV weak and UV upturn galaxies change considerably, which seems to be due to their difference in dispersion. Such dispersion is higher in UV weak systems and it seems to be caused by a wider variety of stellar populations in terms of ages and metallicities.
    
    \item Also, UV upturn galaxies are richer in metals then their UV weak counterparts, which can be seen throughout the different analyses of this paper \citep[which is consistent with the findings of][]{Werle2020}.
    
\end{enumerate}

All the evidences point to the fact that UV weak and upturn galaxies, within the same range of \logm~and $z$, are generally very similar, with few exceptions. The most important differences point to a higher passiveness of UV upturn galaxies and a higher diversity of stellar populations in UV weak systems. Potential explanations to why UV upturn systems have shorter star-formation histories are:

\begin{enumerate}[i.]
    \item if both types of galaxies have been through similar dynamical interactions and other processes, UV upturn galaxies may have simply concluded their star-formation before UV weak galaxies;
    
    \item or, for some reason, UV weak systems have been more efficient in incorporating pristine (metal-poor) gas from the intergalactic medium, stimulating them to create a few more stars than UV upturn galaxies.
\end{enumerate}

Further studies are still required to assess the richness of stellar populations of both these systems as discussed at the end of Sec. \ref{sec:mainresults}, specially by taking into account stellar population templates that are rich in rare stellar evolutionary phases, such as HB, EHB, the entire AGB family of stars (post-AGB, post-early-AGB, and so on), as well as binary systems in interactions.

\section{Data availability} \label{sec:datavailability}
No new data were generated or analysed in support of this research. Instructions on how to acquire the data herein used are described in Sec. \ref{sec:dataset}.

\section*{Acknowledgements}
MLLD acknowledges Coordenação de Aperfeiçoamento de Pessoal de Nível Superior - Brasil (CAPES) - Finance Code 001; and Conselho Nacional de Desenvolvimento Científico e Tecnológico - Brasil (CNPq) project 142294/2018-7. MLLD thanks R.~S.~de Souza for the kind review of the manuscript. MLLD is grateful to Miuchinha for the love and support. PRTC acknowledges financial support from Funda\c{c}\~{a}o de Amparo \`{a} Pesquisa do Estado de S\~{a}o Paulo (FAPESP) process number 2018/05392-8 and CNPq process number 310041/2018-0. We thank the anonymous referee for the suggestions that helped improve this manuscript.

We acknowledge the GAMA team for making their catalogues public and accessible.

The colours used in the figures herewith are colour-blind friendly and were retrieved from \url{www.ColorBrewer.org}. Credits: Cynthia A. Brewer, Geography, Pennsylvania State University.

This work benefited from the following collaborative platforms: \texttt{Overleaf}\footnote{{\url{http://overleaf.com/}}}, \texttt{Github}\footnote{{\url{https://github.com/}}}, \texttt{Slack}\footnote{{\url{https://slack.com}}}, and \texttt{SciServer}\footnote{{\url{http://www.sciserver.org/}}}.

\bibliographystyle{mnras}
\bibliography{paper.bib}

\appendix

\section{Additional plots and tables}

\subsection{Correlations}

\begin{figure*}
    \centering
    \includegraphics[width=0.49\linewidth]{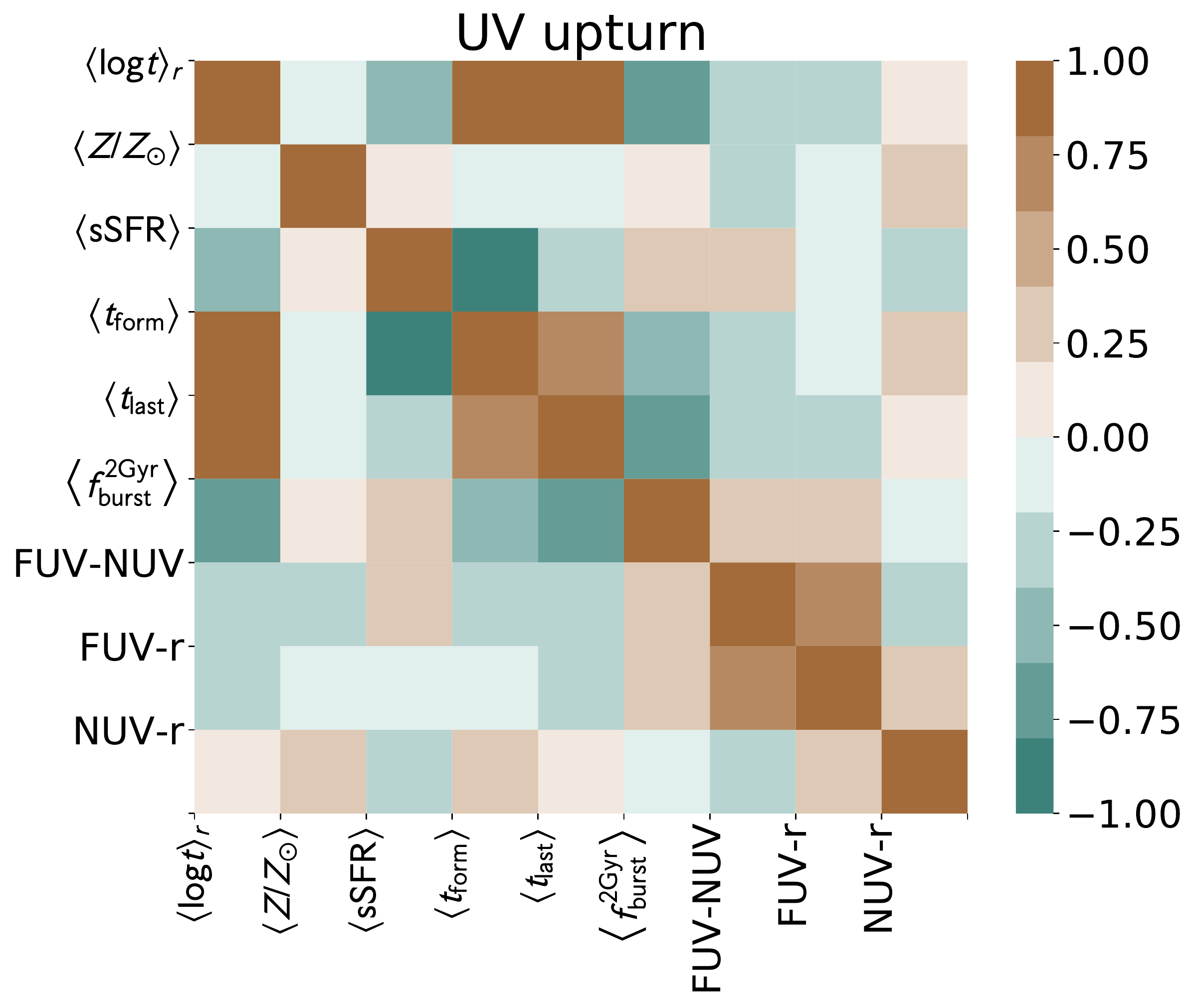}
    \includegraphics[width=0.49\linewidth]{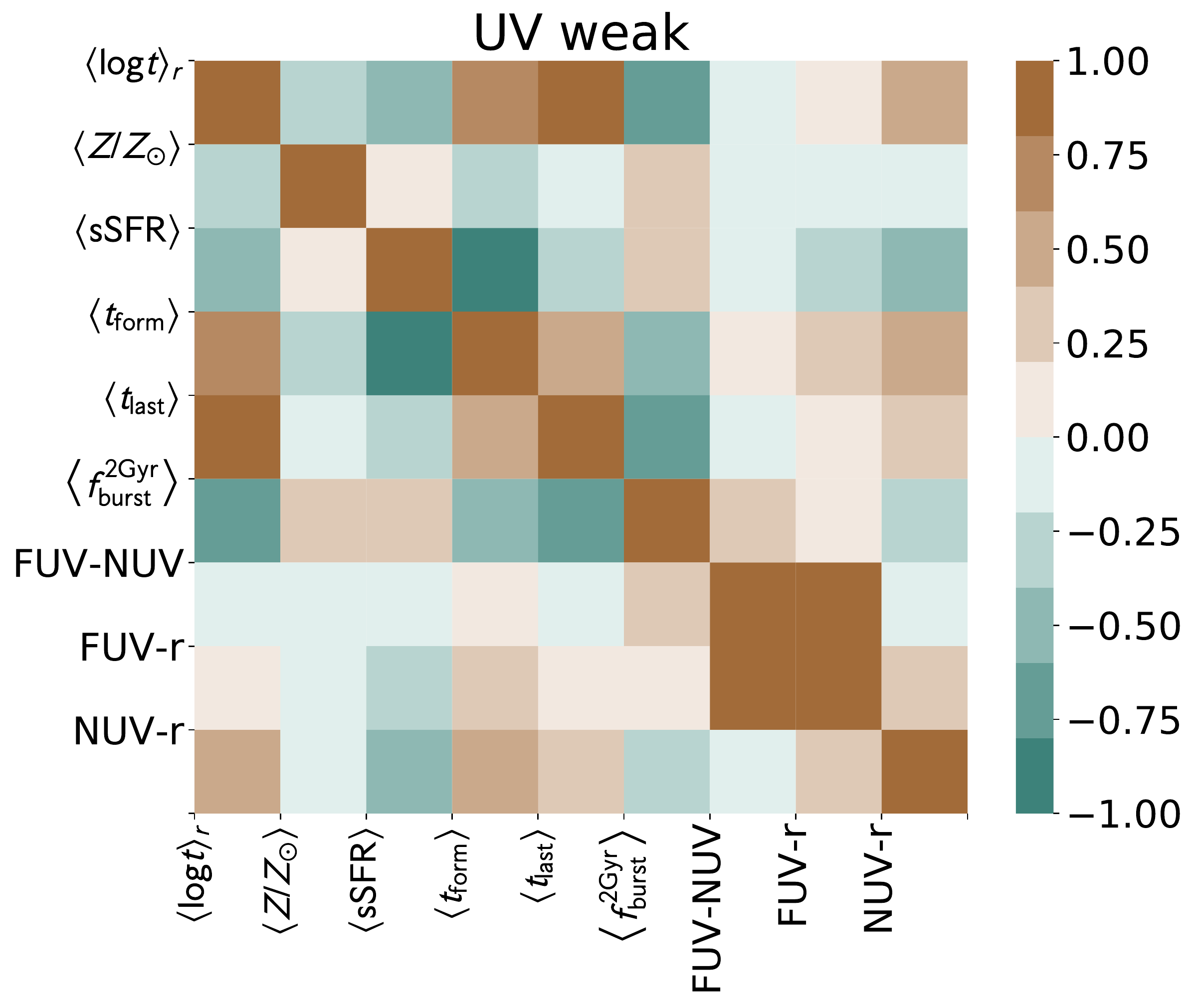}
    \caption{Heatmaps featuring the correlation results ($\rho$, \citealt{Spearman1904}) between \textsc{magphys} resulting parameters. The UV upturn systems are in the left while the UV weak are in the right panel.}
    \label{fig:heatmaps_magphys}
\end{figure*}

We provide an additional image as a visual aid (Fig. \ref{fig:heatmaps_magphys}), as well as tables with the values for the $\rho$ for both UV weak and UV upturn systems (Tables \ref{tab:correlations_uvwk} and \ref{tab:correlations_uvup}). They ease the comparison between the results for both classes of UV bright galaxies, which have been mixed with the clustering effect in Fig. \ref{fig:clustermaps_magphys}.

\begin{table*}
    \centering
    \caption{Correlations table for the UV weak systems.}
    \begin{tabular}{|l|r|r|r|r|r|r|r|r|r|}
    \hline
    & \logt & \metal & \ssfr & \tform & \tlast & \fburst & FUV-NUV & FUV-\rband & NUV-\rband \\ 
    \hline
    \hline
    \logt   &  1.00  & -0.36  & -0.57 	&  0.73  &  0.82  & -0.74  & -0.01  &  0.14  &  0.42 \\ [0.1cm]
    \metal  & -0.36  &  1.00  &  0.12 	& -0.24  & -0.10  &  0.21  & -0.09  & -0.15  & -0.16 \\ [0.1cm]
    \ssfr   & -0.57  &  0.12  &  1.00   & -0.87  & -0.25  &  0.28  & -0.00  & -0.21  & -0.42 \\ [0.1cm]
    \tform  &  0.73  & -0.24  & -0.87 	&  1.00  &  0.42  & -0.40  &  0.01  &  0.20  &  0.42 \\ [0.1cm]
    \tlast  &  0.82  & -0.10  & -0.25 	&  0.42  &	1.00  & -0.76  & -0.06  &  0.08  &  0.36 \\ [0.1cm]
    \fburst	& -0.74  &  0.21  &  0.28 	& -0.40  & -0.76  &  1.00  &  0.32 	&  0.16  & -0.32 \\ [0.1cm]
    FUV-NUV & -0.01  & -0.09  & -0.00 	&  0.01  & -0.06  &  0.32  &  1.00  &  0.82  & -0.14 \\ [0.1cm]
    FUV-\rband  &  0.14  & -0.15  & -0.21 	&  0.20  &	0.08  &  0.16  &  0.82 	&  1.00  &  0.38 \\ [0.1cm]
    NUV-\rband  &  0.42  & -0.16  & -0.42 	&  0.42  &	0.36  & -0.32  & -0.14 	&  0.38  &  1.00 \\ [0.1cm]
    \hline
    \end{tabular}
    \label{tab:correlations_uvwk}
\end{table*}

\begin{table*}
    \centering
    \caption{Correlations table for the UV upturn systems.}
    \begin{tabular}{|l|r|r|r|r|r|r|r|r|r|r|}
    \hline
    &  \logt & \metal & \ssfr & \tform & \tlast & \fburst & FUV-NUV & FUV-\rband & NUV-\rband \\ 
    \hline
    \hline
    \logt   &  1.00  & -0.08  & -0.54  &  0.81  &  0.93  & -0.72  & -0.32  & -0.29  &  0.11 \\ [0.1cm]
    \metal  & -0.08  &  1.00  &  0.08  & -0.06  & -0.00  &  0.03  & -0.24  & -0.05  &  0.30 \\ [0.1cm]
    \ssfr   & -0.54  &  0.08  &  1.00  & -0.83  & -0.31  &  0.40  &  0.22  & -0.02  & -0.37 \\ [0.1cm]
    \tform  &  0.81  & -0.06  & -0.83  &  1.00  &  0.63  & -0.54  & -0.27  & -0.12  &  0.27 \\ [0.1cm]
    \tlast  &  0.93  & -0.00  & -0.31  &  0.63  &  1.00  & -0.74  & -0.32  & -0.33  &  0.05 \\ [0.1cm]
    \fburst & -0.72  &  0.03  &  0.40  & -0.54  & -0.74  &  1.00  &  0.32  &  0.24  & -0.12 \\ [0.1cm]
    FUV-NUV & -0.32  & -0.24  &  0.22  & -0.27  & -0.32  &  0.32  &  1.00  &  0.75  & -0.34 \\ [0.1cm]
    FUV-\rband  & -0.29  & -0.05  & -0.02  & -0.12  & -0.33  &  0.24  &  0.75  &  1.00  &  0.29 \\ [0.1cm]
    NUV-\rband  &  0.11  &  0.30  & -0.37  &  0.27  &  0.05  & -0.12  & -0.34  &  0.29  &  1.00 \\ [0.1cm]
    \hline
    \end{tabular}
    \label{tab:correlations_uvup}
\end{table*}


\bsp	
\label{lastpage}
\end{document}